\documentclass[aps,prl,10pt,twocolumn,superscriptaddress,floatfix,longbibliography]{revtex4-2}

\usepackage{amsmath,amssymb,mathtools,bm}
\usepackage{graphicx}
\usepackage{xcolor}
\usepackage[colorlinks=true,linkcolor=blue!55!black,citecolor=blue!55!black,urlcolor=blue!55!black]{hyperref}

\newcommand{\Jzz}{J_{zz}}
\newcommand{\Jpm}{J_{\pm}}
\newcommand{\Xhat}{\hat{X}}
\newcommand{\OEg}{\mathcal{O}_{E_g}}
\newcommand{\Cov}{\mathrm{Cov}}
\newcommand{\Var}{\mathrm{Var}}
\newcommand{\expv}[1]{\langle #1\rangle}
\newcommand{\Cph}{C_{\mathrm{ring}}}
\newcommand{\geff}{g_{\mathrm{eff}}}
\newcommand{\Gph}{\Gamma_{\mathrm{ring}}}

\makeatletter
\long\def\SM@gobblethree#1#2#3{}
\newcommand{\SMtocoff}{\global\let\SM@addcl\addcontentsline
                       \global\let\addcontentsline\SM@gobblethree}
\newcommand{\SMtocon}{\global\let\addcontentsline\SM@addcl}
\makeatother

\newcommand{\ket}[1]{\lvert #1\rangle}
\newcommand{\bra}[1]{\langle #1\rvert}

\newcommand{\Pice}{\mathcal{P}_0}
\newcommand{\hexagon}{\mathrm{hex}}
\newcommand{\Ppow}{\mathcal{P}}

\begin{document}

\title{Thermodynamic Spectroscopy of Emergent Excitations in Quantum Spin Ice}


\author{Zhengbang Zhou}
\thanks{These authors contributed equally to this work.}
\affiliation{
Department of Physics, University of Toronto, Toronto, Ontario M5S 1A7, Canada
}

\author{Tony An}
\thanks{These authors contributed equally to this work.}
\affiliation{
Department of Physics, University of Toronto, Toronto, Ontario M5S 1A7, Canada
}

\author{Yong Baek Kim}
\affiliation{
 Department of Physics, University of Toronto, Toronto, Ontario M5S 1A7, Canada
}

\date{\today}

\begin{abstract}
Quantum spin ice (QSI) is a three-dimensional quantum spin liquid where fractionalized spinons interact with emergent photons. The XXZ model on the pyrochlore lattice realizes such a $U(1)$ quantum spin liquid and a number of pyrochlore magnets have been investigated as candidate materials, but the detection of emergent excitations has been a major challenge. The specific heat is expected to show the higher-energy spinon excitations and a lower-energy anomaly at the ring-exchange scale, which sets both the photon bandwidth and the energy of the emergent magnetic monopoles (or visons). 
In reality, the lower-energy peak is obscured by the nuclear Schottky anomaly in non-Kramers Pr-based systems, while it is not clearly resolved from the spinon contributions in Ce-based dipolar-octupolar systems. In this work, we propose a novel thermodynamic probe of the elusive ring-exchange energy scale. We show that in the presence of a weak perturbation coupled to the transverse component of the pseudospin degrees of freedom, the temperature derivative of an observable conjugate to such a weak perturbing force is highly sensitive to the ring-exchange energy scale. Using this scheme, it is shown that the difference between thermal expansion coefficients along the [100] and [010] directions should show a peak at the ring-exchange energy scale for non-Kramers QSI. Similarly, the temperature derivative of the magnetization, $dM/dT$, of dipolar–octupolar pyrochlores under a weak magnetic field can also detect the same signal. Moreover, the sign of these signatures distinguishes the zero-flux and $\pi$-flux QSI states.
\end{abstract}

\maketitle

\begin{figure*}[t]
\includegraphics[width=\textwidth]{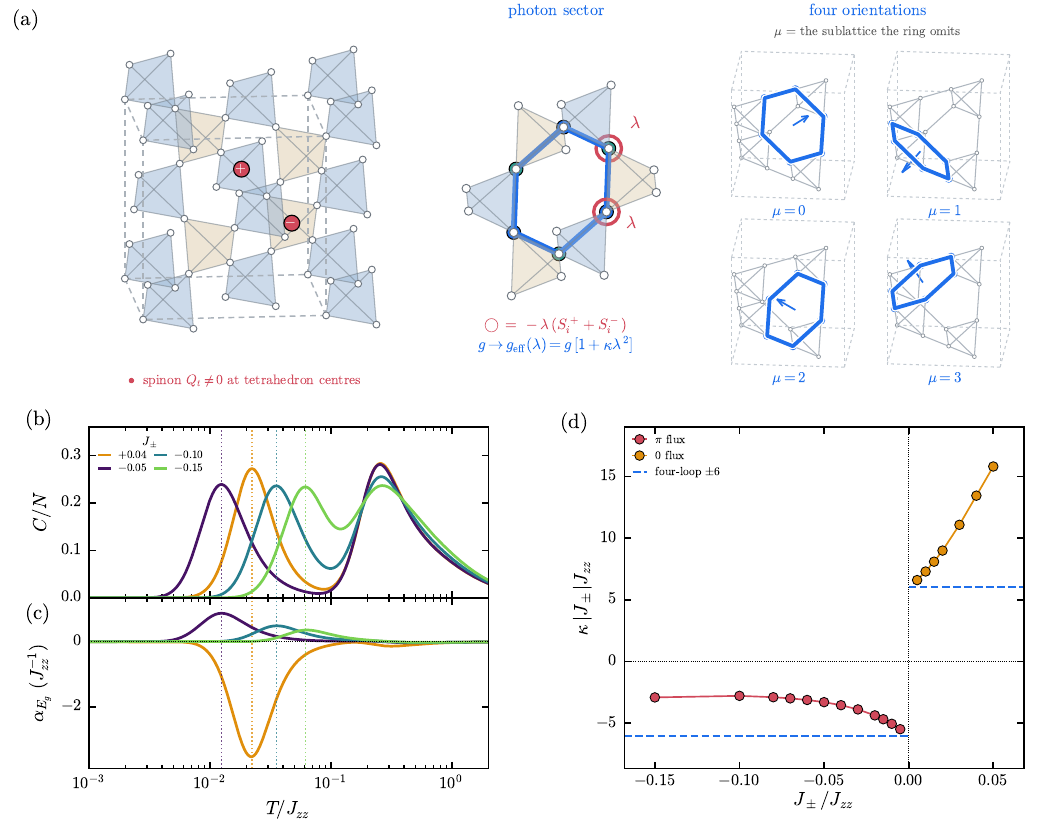}
\caption{\textbf{Anisotropic thermal expansion as ring-exchange spectroscopy in non-Kramers QSI.} (a) Left: Pyrochlore lattice. One spin flip making a spinon pair ($Q_t\neq0$, at tetrahedron centres). Centre: a weak transverse field renormalizes the ring exchange whose virtual process is hosted on a hexagonal ring  [Eq.~\eqref{eq:geff}]. Right: the 16-site cubic cell holds one ring of each orientation, shown separately in a common view; $\mu$ is the sublattice the ring omits, fixing its normal along $\hat{\bm n}_\mu$. (b,c) ED on the 16-site cubic pyrochlore cluster with periodic boundary conditions at $x=\lambda^{2}/(|\Jpm|\Jzz)=0.014$: the $E_g$ response (c) tracks the lower anomaly of $C$ (b), dotted lines, and its sign follows the flux sector, positive at $\pi$ flux and negative at $0$ flux. Couplings from $\Jpm=+0.04$ to $-0.15$ are shown; beyond that the two anomalies of $C$ merge. Note that on this cell the lower scale is the boundary-winding four-loop $g_4=4\Jpm^{2}/\Jzz$ rather than the hexagon $g$~\cite{SM}. (d) Peak-shift coefficient $\kappa|\Jpm|\Jzz=\mathrm{sgn}(\Jpm)\overline{B}(\bm p^{E_g})$ of Eq.~\eqref{eq:kappa}, at $x\in\{0,0.007,0.014,0.021\}$. Both branches approach the four-loop value $\pm6$ appropriate to this cluster (blue dashed) rather than the thermodynamic-limit hexagon value $\pm15$; the sign law is the same for both.}
\label{fig:main}
\end{figure*}

\emph{Introduction.}--- Quantum spin ice (QSI) is a three-dimensional U(1) quantum spin liquid~\cite{Hermele2004,Gingras2014,Rau2019} and its low-energy sector is described by the compact U(1) lattice gauge theory supporting gapped fractionalized spinons and a gapless emergent photon~\cite{Motrunich2002,Moessner2003,Wen2003,Savary2012,Savary2013,Benton2012,Pace2021}. Such a quantum spin liquid can be realized by the nearest-neighbor XXZ model on the pyrochlore lattice,
\begin{equation} H_{\rm XXZ} = \Jzz\sum_{\langle ij\rangle}S_i^zS_j^z -\Jpm\sum_{\langle ij\rangle}\left(S_i^+S_j^-+S_i^-S_j^+\right),
\label{eq:Hxxz}
\end{equation}
written in the local frame of the four pyrochlore sublattices. In the case where $J_\pm=0$, $H_{\rm XXZ}= \frac{J_{zz}}{2}\sum_{t} Q_t^2$, where $ Q_t = \sum_{i\in\delta t} S^z_i$ denotes the spinon charge at some tetrahedron $t$. Therefore, any state with zero spinon charge is a ground state of the system. The resulting large degenerate manifold of spin configurations is dubbed the classical spin ice (CSI)~\cite{Castelnovo2008,Castelnovo2012}. In leading-order perturbation theory for $J_\pm\neq0$, the transverse exchange acting within the ice manifold at third order leads to the ring exchange around a hexagon (see Fig.~\ref{fig:main}(a))~\cite{Hermele2004}. This gives rise to the ring-exchange energy scale $g=12\Jpm^3/\Jzz^{2}$, whose magnitude is the photon bandwidth and whose sign selects the $0$-flux ($\Jpm>0$) or $\pi$-flux ($\Jpm<0$) ground state~\cite{Lee2012,Savary2012,Benton2018Pi,Chern2024,Schaden2024}. The same energy scale also corresponds to emergent magnetic monopoles dubbed visons. Quantum Monte Carlo resolves the two scales as two peaks in the specific heat: a crossover into the CSI near $T \simeq 0.2\Jzz$ associated with the energy of creating a spinon $\sim J_{zz}/2$, and a second, much weaker peak near $T \sim 10^{-3}\Jzz$ below which $C$ acquires the $T^{3}$ signature of a linearly dispersing photon~\cite{Banerjee2008,Shannon2012,Kato2015,Huang2018,Zhou2025QFI}. Only below such a temperature is the emergent photon coherent, and we enter the pristine QSI phase.

Because observation of the lower-temperature peak is closely tied to the identification of QSI, resolving this feature experimentally has been a longstanding goal. Nevertheless, direct access to this lower energy scale has remained challenging despite extensive studies of candidate QSI materials. In the non-Kramers (NK) pyrochlores $\mathrm{Pr_2(Zr,Hf)_2O_7}$, the $^{141}$Pr nucleus has $I=5/2$, and its coupling to the electronic moment produces a hyperfine-enhanced nuclear contribution at low temperatures~\cite{Bleaney1973,Kimura2013,Bonville2016,Gronemann2023}. The corresponding nuclear Schottky entropy gives rise to a large heat-capacity contribution on the $0.1$-K scale that can obscure a much weaker ring-exchange signature~\cite{Blote1969,Kimura2013,Anand2016,Luo2025Pr}. By contrast, in the dipolar--octupolar (DO) cerium pyrochlores $\mathrm{Ce_2(Zr,Hf,Sn)_2O_7}$~\cite{Gaudet2019,Sibille2020,Yahne2024,Poree2024,Yuan2026CeSn}, the naturally abundant Ce isotopes have zero nuclear spin, thereby avoiding this particular Schottky background. In $\mathrm{Ce_2Zr_2O_7}$, thermodynamic and neutron-scattering measurements now provide evidence for low-energy photon-like spectral weight, a $T^{3}$ heat-capacity regime, and a distinct higher-energy spinon continuum~\cite{Gao2019,Smith2022,Gao2025Photons,Gao2026Demarcation}. Even so, a clearly resolved pair of zero-field heat-capacity peaks has not been observed, consistent with the two characteristic energy scales lying relatively close to one another in the experimentally relevant coupling regime~\cite{Smith2022,Smith2023Field,Desrochers2024FiniteT,Gao2025Photons,Zhou2025QFI}. In $\mathrm{Ce_2Hf_2O_7}$, a second low-temperature heat-capacity peak has recently been observed~\cite{Smith2025HfHeat}; however, whether this anomaly corresponds to the QSI crossover itself or instead reflects additional interactions or ordering remains an open question~\cite{Smith2025HfHeat,Poree2025Hf}. These limitations motivate the development of an experimentally accessible thermodynamic probe that is selectively sensitive to the ring-exchange energy scale.

In this Letter, we propose a novel thermodynamic probe of the ring-exchange energy scale. Consider a weak field coupled to the transverse spin components in Eq.~\eqref{eq:Hxxz},
\begin{equation}
H(\lambda)=H_{\rm XXZ}-\lambda\Xhat,
\qquad
\Xhat=\sum_i\left(p_{\mu(i)}S_i^+ + p_{\mu(i)}^*S_i^-\right),
\label{eq:H}
\end{equation}
where $\lambda$ sets the overall strength of the probe and $p_{\mu(i)}\in\mathbb{C}$ specifies its local form factor. Here the form factor depends only on the sublattice: $\mu(i)\in\{0,1,2,3\}$ denotes the sublattice containing site $i$. Throughout, we use a shorthand notation $\bm p=(p_0,p_1,p_2,p_3)$ to denote the form factors on the four sublattices in a pyrochlore lattice. Subscripts on $p_\mu$ always denote sublattice, while the probe channel is indicated by a superscript, as in $\bm p^{E_g}$ below.

Using $\Xhat=-\partial_\lambda H$, we define the thermal response of $\Xhat$ by
\begin{equation}
\gamma_X \equiv \frac{d\expv{\Xhat}}{dT}
= -\frac{\partial^{2}F}{\partial T \partial\lambda}
=\frac{1}{T^{2}}\Cov_T(\Xhat,H),
\label{eq:maxwell}
\end{equation}
where $\Cov_T(A,B)=\langle AB\rangle_T-\langle A\rangle_T\langle B\rangle_T$ denotes the covariance at temperature $T$. We show below that it directly tracks the specific heat, $\gamma_X\propto C(T)$, near the ring-exchange energy scale $g$, 

For NK QSI, the transverse field is realized by strain, which couples to the quadrupolar transverse components $S^\pm$. The thermal response $\gamma_X$ therefore corresponds to a thermal-expansion coefficient, up to constant prefactors. Crucially, the nuclear Schottky background does not contribute directly to this response, allowing the ring-exchange feature to be resolved cleanly.

For DO QSI, consider the regime in which the octupolar component $S_y$ (or $S^x$) is the dominant Ising degree of freedom. The dipolar component $S^z$ serves as the transverse degree of freedom and couples directly to an external magnetic field. In this case, $\gamma_X$ reduces to $d\langle M\rangle/dT$, the temperature derivative of the magnetization. Thus, thermal expansion in NK systems and the temperature derivative of magnetization in DO systems provide experimentally accessible probes of the emergent ring-exchange energy scale.

In what follows, we establish this mechanism perturbatively and verify it using exact diagonalization (ED), computing the specific heat and the corresponding covariances over a range of $\lambda$ and $J_\pm$. We use the 16-site cubic cluster with periodic boundary conditions~\cite{Canals1998,Schafer2020,Wei2023,Wei2024}, diagonalized exactly on the full $2^{16}$-dimensional Hilbert space.

\emph{A transverse field selects the ring-exchange sector.}---Third order degenerate perturbation theory in $\Jpm$ about the CSI manifold produces an effective Hamiltonian consisting of ring-exchange terms acting on the 6-site hexagonal loops of the pyrochlore lattice~\cite{Hermele2004,Sanders2024}. Each hexagon omits one sublattice $\mu$ whose local Ising axis $\hat{\bm n}_\mu$ is normal to the hexagon, so there are four distinct hexagon orientations [Fig.~\ref{fig:main}(a), right]. Collecting each orientation, we obtain the 6-site ring flip operator
\begin{align}
W_\mu=\sum_{h\in\mu}&\left(S^+_1S^-_2S^+_3S^-_4S^+_5S^-_6+\text{h.c.}\right),
\label{eq:Hph}
\end{align}
where $h$ runs over the hexagons of orientation $\mu$ and the sites $1,\dots,6$ traverse the vertices of the hexagon in an arbitrary direction. The effective Hamiltonian is then $H_{\mathrm{ring}}=-g\sum_{\mu}W_\mu$.

When the transverse field $\lambda \Xhat$ is included, the perturbation theory proceeds analogously, but with one of the exchange ($J_\pm$) terms replaced by two field ($\lambda$) insertions~\cite{Sanders2024, SandersYan2024}. The net operation is a hexagon flip with a modified coupling
\begin{align}
\geff^{\mu}&=g\Big[1+B_\mu(\bm p)\frac{\lambda^{2}}{\Jpm\Jzz}+O(\lambda^4)\Big].
\label{eq:geff}
\end{align}
$B_\mu$ is a constant which counts the number of ways the insertion can correct the ring-exchange scale and depends on the form factor ${\bm p}$ corresponding to the relevant perturbation. Moreover, the effective Hamiltonian is renormalized by $\lambda$: $H_{\mathrm{ring}}(\lambda)=-\sum_{\mu}\geff^{\mu}(\lambda)W_\mu$. Since the field reaches the ring-exchange sector only through the couplings $\geff^{\,\mu}$, the chain rule applied to Eq.~\eqref{eq:maxwell} results in
\begin{equation}
\gamma_X\big|_{\mathrm{ring}}
=-\frac{\partial^{2}F}{\partial T\partial\lambda}
=\sum_\mu\frac{\partial\geff^{\mu}}{\partial\lambda}
\frac{\Cov_T(W_\mu,H_{\mathrm{ring}})}{T^{2}}.
\label{eq:chain}
\end{equation}
After some algebra outlined in the SM~\cite{SM}, we arrive at our central result
\begin{align}
    \gamma_{X}\big|_{\mathrm{ring}}&=-\frac{2\overline{B}(\bm p)\lambda}{\Jpm\Jzz}\Cph(T)+O(\lambda^3)\label{eq:lockstep}
\end{align} establishing the leading weak-field proportionality between the thermal response $\gamma_X(T)$ and the specific heat $\Cph=\Cov_T(H_{\mathrm{ring}}, H_{\mathrm{ring}})/T^{2}$. The proportionality constant contains the sublattice-averaged quantities $\overline{B}(\bm p)=\tfrac14\sum_\mu B_\mu(\bm p)$. The relation~\eqref{eq:lockstep} holds at low temperatures $T \ll J_{zz}$ where the effective Hamiltonian \eqref{eq:Hph} is valid; for probes that renormalize all ring orientations equally, it is exact within the one-coupling ring Hamiltonian~\cite{SM}. The position of the ring-exchange peak scales with $|\overline g_{\rm eff}|$ and follows
\begin{equation} T_{\rm peak}(\lambda)=T_{\rm peak}(0)\left[1+\kappa\lambda^{2}+O(\lambda^{4})\right],
\label{eq:kappa}
\end{equation} where we define the shorthand $\kappa=\frac{\overline{B}(\bm p)}{\Jpm\Jzz}$.  Since $\overline{B}(\bm p)\geq0$ for every pattern~\cite{SM}, Eq.~\eqref{eq:lockstep} shows that the sign of $\gamma_X$ is set by the sign of $J_\pm$ alone; for $\pi$-flux QSI, $J_\pm<0$ makes $\gamma_X$ positive, while for 0-flux QSI, $J_\pm>0$ makes $\gamma_X$ negative. In the same light, depending on the sign of $J_\pm$, the transverse field shifts the ring-exchange peak to a lower temperature for $\pi$ flux and to a higher temperature for $0$ flux according to Eq.~\eqref{eq:kappa}. These two signatures combined provide strong diagnostics of the ground-state gauge flux sector.

\emph{Non-Kramers pyrochlores: anisotropic thermal expansion.}---In $\mathrm{Pr_2(Zr,Hf)_2O_7}$, the transverse components of the non-Kramers doublet transform as electric quadrupoles. Consequently, symmetry-breaking strain couples linearly to these components and realizes Eq.~\eqref{eq:H}. Consider the leading order elastic energy and magnetoelastic coupling  $H_{E_g} = \frac{C_{E_g}}{2}\epsilon_{E_g}^2 -g_{E_g}\epsilon_{E_g}\OEg,$ where $g_{E_g}$ is the magnetoelastic coupling, $C_{E_g}$ is the corresponding elastic modulus, and $\epsilon_{E_g}$ is the $E_g$ strain field, then $\lambda=g_{E_g}\epsilon_{E_g}$ can be interpreted as a strain amplitude~\cite{OnodaTanaka2010,Onimaru2016,Patri2020Magnetostriction,Simon2022,Seth2022,Tang2023}. For concreteness, we consider the sublattice-uniform $E_g$ strain couples to 
\begin{equation}
\Xhat=\OEg=\sum_i\big(\sqrt{3}S^x_i-S^y_i\big) =\sum_i e^{i\pi/6}S_i^+ + \mathrm{h.c.}, \label{eq:OEg}
\end{equation} 
with $\bm p^{E_g}=e^{i\pi/6}(1,1,1,1)$, which realizes Eq.~\eqref{eq:lockstep} at $\overline{B}=6$ on a 16-site cubic cluster. The thermal response $\gamma_{E_g}\equiv(\partial_T\langle\OEg\rangle)_\lambda$ is directly proportional to the thermal-expansion coefficient in the corresponding $E_g$ strain channel. To see this, we show that, including a weak applied $E_g$ stress with strength $\sigma_{E_g}$, minimizing over the strain gives, to leading order~\cite{SM}, 
\begin{equation} 
\alpha_{E_g} \equiv\left.\partial_T\epsilon_{E_g}^*\right|_{\sigma_{E_g}} = \frac{g_{E_g}}{C_{E_g}}\gamma_{E_g}. 
\end{equation} More generally the same relation holds with $C_{E_g}$ replaced by the renormalized modulus $C_{E_g}-g_{E_g}^2\chi_{E_g}$, where $\chi_{E_g}=(\partial_\lambda\langle\OEg\rangle)_T$ derived in the SM~\cite{SM}. Thus $\gamma_{E_g}$ is directly probed by anisotropic thermal-expansion measurements up to a prefactor. For simplicity, we use $\alpha_{E_g}$ and $\gamma_{E_g}$ interchangeably.

Experimentally, one can obtain $\alpha_{E_g}$ by taking a difference of thermal expansion coefficients along the cubic $[100]$ and $[010]$ axes~\cite{Patri2019, Patri2020Magnetostriction, SM}
\begin{equation}
    \alpha_{E_g}=\frac{d}{dT}\left(\frac{\Delta L}L\Big|_{[100]} - \frac{\Delta L}L\Big|_{[010]}\right),
    \label{eq:alpha_Eg}
\end{equation}
where $\Delta L/L|_{\hat{\bm \ell}}$ denotes the relative length change along the $\hat{\bm \ell}$ direction. From its definition, $\alpha_{E_g}$ is not sensitive to the nuclear $^{141}$Pr spins. The hyperfine interaction projected onto the non-Kramers doublet is $H_{\rm hf}=A_{\rm hf} \sum_i I^z_i S^z_i$ where $I$ is the nuclear spin~\cite{Bleaney1973,Bonville2016}. This drops out of $\alpha_{E_g} = -\partial_T \partial_\lambda F$ due to the $\lambda$ derivative, so $\alpha_{E_g}$ contains no explicit hyperfine contribution. 

Figures~\ref{fig:main}(b,c) show the computed specific heat $C$ and thermal expansion coefficient $\alpha_{E_g}$, respectively. For all couplings studied, the $\alpha_{E_g}$ response tracks the lower anomaly in $C$ and remains proportional to it through the peak, as shown in Fig~\ref{fig:main}(b,c). The sign of the shift in the peak location is also consistent with $-1/\Jpm$, as required by Eq.~\eqref{eq:lockstep}. Thus, the proportionality relation in Eq.~\eqref{eq:lockstep} is verified nonperturbatively by ED.

Moreover, Figure~\ref{fig:main}(d) shows the extracted quantity $\kappa|\Jpm|\Jzz=\mathrm{sgn}(\Jpm)\overline{B}(\bm p^{E_g})$. As $|\Jpm|\to0$, it approaches the precise factor $\pm 6 = \pm \overline B(\bm p^{E_g})$ obtained by perturbation theory. The agreement in the perturbative limit confirms that the renormalization enters only as a correction to the ring-exchange term. We further emphasize that the sign of $\kappa$ provides a direct diagnostic of the underlying flux sector. Indeed, as shown in Fig.~\ref{fig:main}(d), in the $\pi$-flux regime, where $J_\pm<0$, increasing the field shifts the ring-exchange scale downward, whereas in the 0-flux regime it shifts the scale upward.

\begin{figure}[t]
\includegraphics[width=\linewidth]{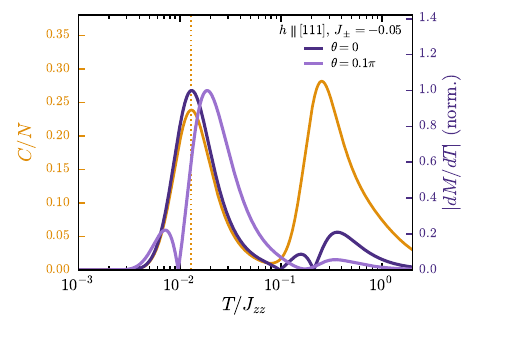}
\caption{\textbf{The mixing angle sets where the magnetocaloric response sits.} ED on the 16-site cubic cluster at $\Jpm=-0.05$ with a $[111]$ field [Eq.~\eqref{eq:HZ}] with field strength $h^2=0.014 |J_\pm| J_{zz}$. For the purely transverse coupling $\theta=0$ the extremum of $|dM/dT|$ (blue) sits on the lower peak of $C$ (black, dotted line); for $\theta\neq0$ ($\theta=0.1\pi$, orange) the field lifts the ice degeneracy at first order and the response moves onto the field-set scale on which the ice manifold reorganizes, a factor $12$ above the hexagon scale $g$ although only a factor $1.4$ above this cluster's own $g_4$. Both $|dM/dT|$ curves are normalized to their maxima.}
\label{fig:do}
\end{figure}

\emph{Dipolar--octupolar pyrochlores: magnetocaloric response.}---The distinctive field response of dipolar--octupolar (DO) pyrochlores follows directly from the symmetry of the Kramers doublet. In the pseudospin basis, $\tau^x$ and $\tau^z$ transform identically as magnetic dipoles under the point group, whereas $\tau^y$ transforms as a magnetic octupole~\cite{Huang2014,Li2017}. Therefore, the exchange Hamiltonian generically contains a mixed term $J_{xz}\tau_i^x\tau_j^z+\text{h.c.}$. This mixing can be removed by a rotation in the $(\tau^x,\tau^z)$ plane, $\tau^z=S^z\cos\theta+S^x\sin\theta,$ with the material-dependent angle $\theta$ set by the microscopic exchange couplings~\cite{Huang2014,Gaudet2019,Smith2022,Smith2023Field,Zhou2024DOfield}.

This distinction becomes important in a magnetic field. The Zeeman coupling only acts on the moment $\tau^z$, and therefore, in the rotated exchange basis, couples to the combination $S^z\cos\theta+S^x\sin\theta$~\cite{Huang2014,Li2017,Desrochers2022,Zhou2024DOfield,Zhou2025PhaseDiagram}. Consider the $S^{x}$ as the dominant Ising degree of freedom relevant to the Ce-based pyrochlores according to thermodynamic and neutron-scattering analyses~\cite{Sibille2020,Smith2022,Bhardwaj2022,Smith2025Octupolar}. Here, we take $S^x$ to be the Ising variable carrying the gauge charge $Q_t$, with $S^{y,z}$ transverse. The perturbation $\lambda\hat X$ can then be identified with the Zeeman coupling, with $\lambda=h$ and
\begin{equation}
\hat X=M=\sum_i(\hat{\bm n}_i\cdot\hat{\bm h})
\left(S_i^z\cos\theta+S_i^x\sin\theta\right),
\label{eq:HZ}
\end{equation}
where $\hat{\bm n}_i$ is the local easy axis and $\hat{\bm h}$ specifies the field direction. Since $M$ is conjugate to $h$, Eq.~\eqref{eq:maxwell} applies directly with $\gamma_X\rightarrow dM/dT$. The remaining spatial form factor is fixed by the field orientation,
$p_\mu=\hat{\bm n}_\mu\cdot\hat{\bm h}$. For the $[111]$ field considered throughout,
$\bm p^{[111]}=\left(1,-\tfrac13,-\tfrac13,-\tfrac13\right)$.

The mixing angle $\theta$ describes how much of the Zeeman coupling is transverse to the ice manifold. At $\theta=0$, the field couples only to $S^z$ and is therefore purely transverse. Consequently, $dM/dT$ obeys the same proportionality with the specific heat in Eq.~\eqref{eq:lockstep}. Figure~\ref{fig:do} confirms that $|dM/dT|$ peaks at the lower anomaly of $C$. For $\theta\neq0$, however, the field also couples to $S^x$ and acts directly within the ice manifold, lifting the degeneracy at first order in $h$. The dominant response is then parametrically unrelated to $g\propto|\Jpm|^3$: in the example of Fig.~\ref{fig:do} it sits a factor of 12 above $g$. The position of the low-temperature $|dM/dT|$ extremum relative to the ring-exchange energy scale is therefore sensitive to the mixing angle $\theta$. Existing thermodynamic and neutron analyses favor a small mixing angle in $\mathrm{Ce_2Zr_2O_7}$, but the inferred value remains model- and dataset-dependent~\cite{Bhardwaj2022,Desrochers2022,Smith2023Field,Zhou2024DOfield,Smith2025Octupolar,Zhou2025PhaseDiagram,Gao2026Demarcation}, so this particular measurement would give a sharp diagnostic.

\begin{figure}[t]
\includegraphics[width=\linewidth]{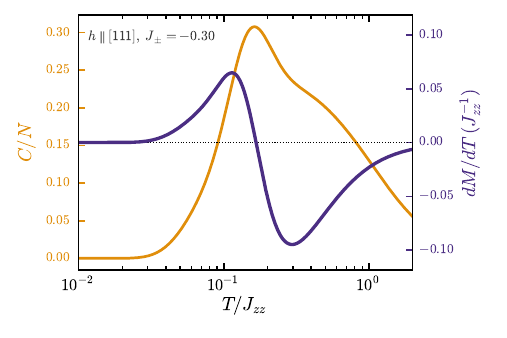}
\caption{\textbf{Specific heat and $dM/dT$ at $J_\pm=-0.30$ with field strength $h^2=0.014 |J_\pm| J_{zz}$ and $\theta=0$} Here, the two energy scales are close enough that $C$ (black) has a single merged maximum. The signed $dM/dT$ (blue) can still resolve the two energy scales: positive at the ring-exchange scale and negative at $T\sim\Jzz$.}
\label{fig:do30}
\end{figure}

The same measurement reveals something that the specific heat cannot. Figure~\ref{fig:do30} shows the case $\Jpm=-0.30$, where the two energy scales have moved close enough together that the two-peak structure expected in $C(T)$ has collapsed into a single broad peak. In contrast, $dM/dT$ can still distinguish the two scales: a positive lobe at the lower, ring-exchange energy scale and a negative dip at $T\sim\Jzz/2$, associated with the spinon energy scale. The key difference is the relative sign of the two contributions. In the specific heat, the spinon and ring-exchange features contribute with the same sign and therefore merge as their characteristic temperatures approach one another. In $dM/dT$, however, the spinon response is negative while the ring-exchange response is positive. As a result, even at large $J_\pm$, where the two energy scales would otherwise be difficult to distinguish, $dM/dT$ can continue to resolve them through the opposite signs of their responses. A detailed discussion of why the spinon signature is negative is given in the SM~\cite{SM}. For the case $\hat{\bm h}\parallel[111]$ considered here, the sign has a simple physical interpretation. At intermediate temperatures, thermally excited spinons—3-in-1-out defects of the 2-in-2-out ice rule—begin to proliferate. The $[111]$ field preferentially selects those defects whose net moment lies along the field, producing a finite magnetization $M$. As the temperature is raised further, spinons of all orientations become thermally populated, reducing the net magnetization. This crossover occurs at a temperature set by the spinon gap, of order $J_{zz}/2$, and gives rise to the negative spinon feature in $dM/dT$, in contrast to the positive ring-exchange response.

\emph{Outlook.}---
For the NK QSI candidates $\mathrm{Pr_2(Zr,Hf)_2O_7}$, we find that the anisotropic thermal expansion $\alpha_{E_g}$ provides a clean probe of the ring-exchange scale, free from the direct nuclear hyperfine contribution that can obscure the corresponding signature in the specific heat. Moreover, the direction of the peak shift under increasing strain distinguishes the ground-state flux sector: the $\pi$-flux peak shifts to lower temperatures, whereas the $0$-flux peak shifts to higher temperatures. The same framework extends naturally to the DO QSI candidates $\mathrm{Ce_2(Zr,Hf,Sn)_2O_7}$, where, for dominant Ising $S^x$ (or $S^y$), the magnetocaloric response $dM/dT$ plays an analogous role. In addition to tracking the ring-exchange scale, $dM/dT$ can help disentangle the energy scales of ring-exchange and spinon features that may merge in the specific heat, while also providing a sensitive diagnostic of a finite mixing angle $\theta$.

Taken together, our results establish covariance-based thermodynamic responses as powerful probes of the low-energy structure of QSI. By choosing perturbations that couple selectively to the transverse pseudospin components, one can isolate signatures of ring-exchange energy scale that determines the emergent photon bandwidth and vison energy.
Such signatures are otherwise obscured in conventional thermodynamic observables. These measurements thus provide a practical route to isolating the ring-exchange scale and extracting information about the low-energy effective description of candidate QSI materials that is otherwise difficult to obtain.

\SMtocoff
\begin{acknowledgments}
We thank Romain Sibille and Ilaria Villa for discussions and collaboration in a related project, which motivated this work.
T.A., Z.Z., and Y.B.K.
were supported by the Natural Sciences and Engineering Research Council of Canada (NSERC) Grant No. RGPIN-2023-03296 and the Centre for Quantum Materials at the University of Toronto. Computations at the University of Toronto were performed on the Fir and the Rorqual cluster, which is hosted by the Digital Research Alliance of Canada. T.A. and Z.Z. are further supported by the Ontario Graduate Scholarship. 
\end{acknowledgments}

\bibliography{refs}

@article{Hermele2004,
  author  = {Hermele, Michael and Fisher, Matthew P. A. and Balents, Leon},
  title   = {Pyrochlore photons: The {{U(1)}} spin liquid in a {$S=\frac{1}{2}$} three-dimensional frustrated magnet},
  journal = {Phys. Rev. B},
  volume  = {69},
  pages   = {064404},
  year    = {2004},
  doi     = {10.1103/PhysRevB.69.064404}
}

@article{Gingras2014,
  author  = {Gingras, M. J. P. and McClarty, P. A.},
  title   = {Quantum spin ice: a search for gapless quantum spin liquids in pyrochlore magnets},
  journal = {Rep. Prog. Phys.},
  volume  = {77},
  pages   = {056501},
  year    = {2014},
  doi     = {10.1088/0034-4885/77/5/056501}
}

@article{Savary2012,
  author  = {Savary, Lucile and Balents, Leon},
  title   = {Coulombic Quantum Liquids in Spin-1/2 Pyrochlores},
  journal = {Phys. Rev. Lett.},
  volume  = {108},
  pages   = {037202},
  year    = {2012},
  doi     = {10.1103/PhysRevLett.108.037202}
}

@article{Benton2012,
  author  = {Benton, Owen and Sikora, Olga and Shannon, Nic},
  title   = {Seeing the light: Experimental signatures of emergent electromagnetism in a quantum spin ice},
  journal = {Phys. Rev. B},
  volume  = {86},
  pages   = {075154},
  year    = {2012},
  doi     = {10.1103/PhysRevB.86.075154}
}

@article{Rau2019,
  author  = {Rau, Jeffrey G. and Gingras, Michel J. P.},
  title   = {Frustrated Quantum Rare-Earth Pyrochlores},
  journal = {Annu. Rev. Condens. Matter Phys.},
  volume  = {10},
  pages   = {357},
  year    = {2019},
  doi     = {10.1146/annurev-conmatphys-022317-110520}
}

@article{Ross2011,
  author  = {Ross, Kate A. and Savary, Lucile and Gaulin, Bruce D. and Balents, Leon},
  title   = {Quantum Excitations in Quantum Spin Ice},
  journal = {Phys. Rev. X},
  volume  = {1},
  pages   = {021002},
  year    = {2011},
  doi     = {10.1103/PhysRevX.1.021002}
}

@article{Kimura2013,
  author  = {Kimura, K. and Nakatsuji, S. and Wen, J.-J. and Broholm, C. and Stone, M. B. and Nishibori, E. and Sawa, H.},
  title   = {Quantum fluctuations in spin-ice-like {{Pr$_2$Zr$_2$O$_7$}}},
  journal = {Nat. Commun.},
  volume  = {4},
  pages   = {1934},
  year    = {2013},
  doi     = {10.1038/ncomms2914}
}

@article{Bonville2016,
  author  = {Bonville, P. and Guitteny, S. and Gukasov, A. and Mirebeau, I. and Petit, S. and Decorse, C. and Ciomaga Hatnean, M. and Balakrishnan, G.},
  title   = {Magnetic properties and crystal field in {{Pr$_2$Zr$_2$O$_7$}}},
  journal = {Phys. Rev. B},
  volume  = {94},
  pages   = {134428},
  year    = {2016},
  doi     = {10.1103/PhysRevB.94.134428}
}

@article{Anand2016,
  author  = {Anand, V. K. and Opherden, L. and Xu, J. and Adroja, D. T. and Islam, A. T. M. N. and Herrmannsd{\"o}rfer, T. and Hornung, J. and Sch{\"o}nemann, R. and Uhlarz, M. and Walker, H. C. and Casati, N. and Lake, B.},
  title   = {Physical properties of the candidate quantum spin-ice system {{Pr$_2$Hf$_2$O$_7$}}},
  journal = {Phys. Rev. B},
  volume  = {94},
  pages   = {144415},
  year    = {2016},
  doi     = {10.1103/PhysRevB.94.144415}
}

@article{Gaudet2019,
  author  = {Gaudet, J. and Smith, E. M. and Dudemaine, J. and Beare, J. and Buhariwalla, C. R. C. and Butch, N. P. and Stone, M. B. and Kolesnikov, A. I. and Xu, Guangyong and Yahne, D. R. and others},
  title   = {Quantum Spin Ice Dynamics in the Dipole-Octupole Pyrochlore Magnet {{Ce$_2$Zr$_2$O$_7$}}},
  journal = {Phys. Rev. Lett.},
  volume  = {122},
  pages   = {187201},
  year    = {2019},
  doi     = {10.1103/PhysRevLett.122.187201}
}

@article{Smith2022,
  author  = {Smith, E. M. and Benton, O. and Yahne, D. R. and Placke, B. and Sch{\"a}fer, R. and Gaudet, J. and Dudemaine, J. and Fitterman, A. and Beare, J. and Wildes, A. R. and others},
  title   = {Case for a {{U(1)$_\pi$}} Quantum Spin Liquid Ground State in the Dipole-Octupole Pyrochlore {{Ce$_2$Zr$_2$O$_7$}}},
  journal = {Phys. Rev. X},
  volume  = {12},
  pages   = {021015},
  year    = {2022},
  doi     = {10.1103/PhysRevX.12.021015}
}

@article{Smith2023Field,
  author  = {Smith, E. M. and Dudemaine, J. and Placke, B. and Sch{\"a}fer, R. and Yahne, D. R. and DeLazzer, T. and Fitterman, A. and Beare, J. and Gaudet, J. and Buhariwalla, C. R. C. and Podlesnyak, A. and Xu, Guangyong and Clancy, J. P. and Movshovich, R. and Luke, G. M. and Ross, K. A. and Moessner, R. and Benton, O. and Bianchi, A. D. and Gaulin, B. D.},
  title   = {Quantum Spin Ice Response to a Magnetic Field in the Dipole-Octupole Pyrochlore {{Ce$_2$Zr$_2$O$_7$}}},
  journal = {Phys. Rev. B},
  volume  = {108},
  pages   = {054438},
  year    = {2023},
  doi     = {10.1103/PhysRevB.108.054438}
}

@article{Gao2019,
  author  = {Gao, Bin and Chen, Tong and Tam, David W. and Huang, Chien-Lung and Sasmal, Kalyan and Adroja, Devashibhai T. and Ye, Feng and Cao, Huibo and Sala, Gabriele and Stone, Matthew B. and others},
  title   = {Experimental signatures of a three-dimensional quantum spin liquid in effective spin-1/2 {{Ce$_2$Zr$_2$O$_7$}} pyrochlore},
  journal = {Nat. Phys.},
  volume  = {15},
  pages   = {1052},
  year    = {2019},
  doi     = {10.1038/s41567-019-0577-6}
}

@article{Petit2016,
  author  = {Petit, S. and Lhotel, E. and Canals, B. and Ciomaga Hatnean, M. and Ollivier, J. and Mutka, H. and Ressouche, E. and Wildes, A. R. and Lees, M. R. and Balakrishnan, G.},
  title   = {Observation of magnetic fragmentation in spin ice},
  journal = {Nat. Phys.},
  volume  = {12},
  pages   = {746},
  year    = {2016},
  doi     = {10.1038/nphys3710}
}

@article{Poree2024,
  author  = {Por{\'e}e, Victor and Yan, Han and Desrochers, F{\'e}lix and Petit, Sylvain and Lhotel, Elsa and Appel, Markus and Ollivier, Jacques and Kim, Yong Baek and Nevidomskyy, Andriy H. and Sibille, Romain},
  title   = {Evidence for fractional matter coupled to an emergent gauge field in a quantum spin ice},
  journal = {Nat. Phys.},
  volume  = {21},
  pages   = {83},
  year    = {2025},
  doi     = {10.1038/s41567-024-02711-w}
}

@article{Smith2025HfHeat,
  author  = {Smith, E. M. and Fitterman, A. and Sch{\"a}fer, R. and Placke, B. and Woods, A. and Lee, S. and Huang, S. H.-Y. and Beare, J. and Sharma, S. and Chatterjee, D. and Balz, C. and Stone, M. B. and Kolesnikov, A. I. and Wildes, A. R. and Kermarrec, E. and Luke, G. M. and Benton, O. and Moessner, R. and Movshovich, R. and Bianchi, A. D. and Gaulin, B. D.},
  title   = {Two-Peak Heat Capacity Accounts for {$R\ln(2)$} Entropy and Ground State Access in the Dipole-Octupole Pyrochlore {{Ce$_2$Hf$_2$O$_7$}}},
  journal = {Phys. Rev. Lett.},
  volume  = {135},
  pages   = {086702},
  year    = {2025},
  doi     = {10.1103/4qxy-l8pg}
}

@article{Poree2025Hf,
  author  = {Por{\'e}e, Victor and Bhardwaj, Anish and Lhotel, Elsa and Petit, Sylvain and Gauthier, Nicolas and Yan, Han and Pomjakushin, Vladimir and Ollivier, Jacques and Quilliam, Jeffrey A. and Nevidomskyy, Andriy H. and others},
  title   = {Dipolar-octupolar correlations and hierarchy of exchange interactions in {{Ce$_2$Hf$_2$O$_7$}}},
  journal = {Phys. Rev. B},
  volume  = {112},
  pages   = {L180404},
  year    = {2025},
  doi     = {10.1103/j451-ztvr}
}

@article{Yahne2024,
  author  = {Yahne, D. R. and Placke, B. and Sch{\"a}fer, R. and Benton, O. and Moessner, R. and Powell, M. and Kolis, J. W. and Pasco, C. M. and May, A. F. and Frontzek, M. D. and others},
  title   = {Dipolar Spin Ice Regime Proximate to an All-In-All-Out {{N\'eel}} Ground State in the Dipolar-Octupolar Pyrochlore {{Ce$_2$Sn$_2$O$_7$}}},
  journal = {Phys. Rev. X},
  volume  = {14},
  pages   = {011005},
  year    = {2024},
  doi     = {10.1103/PhysRevX.14.011005}
}

@article{Patri2019,
  author  = {Patri, Adarsh S. and Hosoi, Masashi and Kim, Yong Baek},
  title   = {Distinguishing dipolar and octupolar quantum spin ices using contrasting magnetostriction signatures},
  journal = {Phys. Rev. Research},
  volume  = {2},
  pages   = {023253},
  year    = {2020},
  doi     = {10.1103/PhysRevResearch.2.023253}
}

@article{Huang2014,
  author  = {Huang, Yi-Ping and Chen, Gang and Hermele, Michael},
  title   = {Quantum Spin Ices and Topological Phases from Dipolar-Octupolar Doublets on the Pyrochlore Lattice},
  journal = {Phys. Rev. Lett.},
  volume  = {112},
  pages   = {167203},
  year    = {2014},
  doi     = {10.1103/PhysRevLett.112.167203}
}

@article{Li2017,
  author  = {Li, Yao-Dong and Chen, Gang},
  title   = {Symmetry enriched {{U(1)}} topological orders for dipole-octupole doublets on a pyrochlore lattice},
  journal = {Phys. Rev. B},
  volume  = {95},
  pages   = {041106},
  year    = {2017},
  doi     = {10.1103/PhysRevB.95.041106}
}

@article{Chen2017,
  author  = {Chen, Gang},
  title   = {Spectral periodicity of the spinon continuum in quantum spin ice},
  journal = {Phys. Rev. B},
  volume  = {96},
  pages   = {085136},
  year    = {2017},
  doi     = {10.1103/PhysRevB.96.085136}
}

@article{Desrochers2022,
  author  = {Desrochers, F{\'e}lix and Chern, Li Ern and Kim, Yong Baek},
  title   = {Competing {{U(1)}} and {$\mathbb{Z}_2$} dipolar-octupolar quantum spin liquids on the pyrochlore lattice: Application to {{Ce$_2$Zr$_2$O$_7$}}},
  journal = {Phys. Rev. B},
  volume  = {105},
  pages   = {035149},
  year    = {2022},
  doi     = {10.1103/PhysRevB.105.035149}
}

@article{Lee2012,
  author  = {Lee, SungBin and Onoda, Shigeki and Balents, Leon},
  title   = {Generic quantum spin ice},
  journal = {Phys. Rev. B},
  volume  = {86},
  pages   = {104412},
  year    = {2012},
  doi     = {10.1103/PhysRevB.86.104412}
}

@article{Benton2018Pi,
  author  = {Benton, Owen and Jaubert, L. D. C. and Singh, Rajiv R. P. and Oitmaa, Jaan and Shannon, Nic},
  title   = {Quantum Spin Ice with Frustrated Transverse Exchange: From a {$\pi$}-Flux Phase to a Nematic Quantum Spin Liquid},
  journal = {Phys. Rev. Lett.},
  volume  = {121},
  pages   = {067201},
  year    = {2018},
  doi     = {10.1103/PhysRevLett.121.067201}
}

@article{Bhardwaj2022,
  author  = {Bhardwaj, Anish and Zhang, Shu and Yan, Han and Moessner, Roderich and Nevidomskyy, Andriy H. and Changlani, Hitesh J.},
  title   = {Sleuthing out exotic quantum spin liquidity in the pyrochlore magnet {{Ce$_2$Zr$_2$O$_7$}}},
  journal = {npj Quantum Mater.},
  volume  = {7},
  pages   = {51},
  year    = {2022},
  doi     = {10.1038/s41535-022-00458-2}
}

@article{Banerjee2008,
  author  = {Banerjee, Argha and Isakov, Sergei V. and Damle, Kedar and Kim, Yong Baek},
  title   = {Unusual Liquid State of Hard-Core Bosons on the Pyrochlore Lattice},
  journal = {Phys. Rev. Lett.},
  volume  = {100},
  pages   = {047208},
  year    = {2008},
  doi     = {10.1103/PhysRevLett.100.047208}
}

@article{Shannon2012,
  author  = {Shannon, Nic and Sikora, Olga and Pollmann, Frank and Penc, Karlo and Fulde, Peter},
  title   = {Quantum Ice: A Quantum {{Monte Carlo}} Study},
  journal = {Phys. Rev. Lett.},
  volume  = {108},
  pages   = {067204},
  year    = {2012},
  doi     = {10.1103/PhysRevLett.108.067204}
}

@article{Kato2015,
  author  = {Kato, Yasuyuki and Onoda, Shigeki},
  title   = {Numerical Evidence of Quantum Melting of Spin Ice: Quantum-to-Classical Crossover},
  journal = {Phys. Rev. Lett.},
  volume  = {115},
  pages   = {077202},
  year    = {2015},
  doi     = {10.1103/PhysRevLett.115.077202}
}

@article{Huang2018,
  author  = {Huang, Chun-Jiong and Deng, Youjin and Wan, Yuan and Meng, Zi Yang},
  title   = {Dynamics of Topological Excitations in a Model Quantum Spin Ice},
  journal = {Phys. Rev. Lett.},
  volume  = {120},
  pages   = {167202},
  year    = {2018},
  doi     = {10.1103/PhysRevLett.120.167202}
}

@article{Savary2013,
  author  = {Savary, Lucile and Balents, Leon},
  title   = {Spin liquid regimes at nonzero temperature in quantum spin ice},
  journal = {Phys. Rev. B},
  volume  = {87},
  pages   = {205130},
  year    = {2013},
  doi     = {10.1103/PhysRevB.87.205130}
}

@article{Chen2016,
  author  = {Chen, Gang},
  title   = {``{{Magnetic}} monopole'' condensation of the pyrochlore ice {{U(1)}} quantum spin liquid: Application to {{Pr$_2$Ir$_2$O$_7$}} and {{Yb$_2$Ti$_2$O$_7$}}},
  journal = {Phys. Rev. B},
  volume  = {94},
  pages   = {205107},
  year    = {2016},
  doi     = {10.1103/PhysRevB.94.205107}
}

@article{Desrochers2023,
  author  = {Desrochers, F{\'e}lix and Chern, Li Ern and Kim, Yong Baek},
  title   = {Symmetry fractionalization in the gauge mean-field theory of quantum spin ice},
  journal = {Phys. Rev. B},
  volume  = {107},
  pages   = {064404},
  year    = {2023},
  doi     = {10.1103/PhysRevB.107.064404}
}

@article{Desrochers2024,
  author  = {Desrochers, F{\'e}lix and Kim, Yong Baek},
  title   = {Spectroscopic Signatures of Fractionalization in Octupolar Quantum Spin Ice},
  journal = {Phys. Rev. Lett.},
  volume  = {132},
  pages   = {066502},
  year    = {2024},
  doi     = {10.1103/PhysRevLett.132.066502}
}

@article{Motrunich2002,
  author  = {Motrunich, O. I. and Senthil, T.},
  title   = {Exotic Order in Simple Models of Bosonic Systems},
  journal = {Phys. Rev. Lett.},
  volume  = {89},
  pages   = {277004},
  year    = {2002},
  doi     = {10.1103/PhysRevLett.89.277004}
}

@article{Moessner2003,
  author  = {Moessner, R. and Sondhi, S. L.},
  title   = {Three-dimensional resonating-valence-bond liquids and their excitations},
  journal = {Phys. Rev. B},
  volume  = {68},
  pages   = {184512},
  year    = {2003},
  doi     = {10.1103/PhysRevB.68.184512}
}

@article{Wen2003,
  author  = {Wen, Xiao-Gang},
  title   = {Artificial light and quantum order in systems of screened dipoles},
  journal = {Phys. Rev. B},
  volume  = {68},
  pages   = {115413},
  year    = {2003},
  doi     = {10.1103/PhysRevB.68.115413}
}

@misc{SandersYan2024,
  title={Experimentally tunable QED in dipolar-octupolar quantum spin ice}, 
  author={Alaric Sanders and Han Yan and Claudio Castelnovo and Andriy H. Nevidomskyy},
  year={2024},
  eprint={2312.11641},
  archivePrefix={arXiv},
  primaryClass={cond-mat.str-el},
  url={https://arxiv.org/abs/2312.11641}, 
}

@article{Sanders2024,
  title = {Vison crystal in quantum spin ice on the breathing pyrochlore lattice},
  author = {Sanders, Alaric and Castelnovo, Claudio},
  journal = {Phys. Rev. B},
  volume = {109},
  issue = {9},
  pages = {094426},
  numpages = {24},
  year = {2024},
  month = {Mar},
  publisher = {American Physical Society},
  doi = {10.1103/PhysRevB.109.094426},
  url = {https://link.aps.org/doi/10.1103/PhysRevB.109.094426}
}

@article{Jaubert2009,
    author = {Jaubert, L. and Holdsworth, P.},
    title = {Signature of magnetic monopole and {D}irac string dynamics in spin ice},
    journal = {Nature Physics},
    volume  = {5},
    pages   = {258},
    year    = {2009},
    doi     = {10.1038/nphys1227}  
}

@article{Castelnovo2008,
  author  = {Castelnovo, C. and Moessner, R. and Sondhi, S. L.},
  title   = {Magnetic monopoles in spin ice},
  journal = {Nature},
  volume  = {451},
  pages   = {42},
  year    = {2008},
  doi     = {10.1038/nature06433}
}

@article{Fennell2009,
  author  = {Fennell, T. and Deen, P. P. and Wildes, A. R. and Schmalzl, K. and Prabhakaran, D. and Boothroyd, A. T. and Aldus, R. J. and McMorrow, D. F. and Bramwell, S. T.},
  title   = {Magnetic {{Coulomb}} Phase in the Spin Ice {{Ho$_2$Ti$_2$O$_7$}}},
  journal = {Science},
  volume  = {326},
  pages   = {415},
  year    = {2009},
  doi     = {10.1126/science.1177582}
}

@article{Morris2009,
  author  = {Morris, D. J. P. and Tennant, D. A. and Grigera, S. A. and Klemke, B. and Castelnovo, C. and Moessner, R. and Czternasty, C. and Meissner, M. and Rule, K. C. and Hoffmann, J.-U. and others},
  title   = {{{Dirac}} Strings and Magnetic Monopoles in the Spin Ice {{Dy$_2$Ti$_2$O$_7$}}},
  journal = {Science},
  volume  = {326},
  pages   = {411},
  year    = {2009},
  doi     = {10.1126/science.1178868}
}

@article{Castelnovo2012,
  author  = {Castelnovo, C. and Moessner, R. and Sondhi, S. L.},
  title   = {Spin Ice, Fractionalization, and Topological Order},
  journal = {Annu. Rev. Condens. Matter Phys.},
  volume  = {3},
  pages   = {35},
  year    = {2012},
  doi     = {10.1146/annurev-conmatphys-020911-125058}
}

@article{Patri2020Magnetostriction,
  author  = {Patri, Adarsh S. and Hosoi, Masashi and Lee, SungBin and Kim, Yong Baek},
  title   = {Theory of magnetostriction for multipolar quantum spin ice in pyrochlore materials},
  journal = {Phys. Rev. Research},
  volume  = {2},
  pages   = {033015},
  year    = {2020},
  doi     = {10.1103/PhysRevResearch.2.033015}
}

@article{Patri2019Unveiling,
    author = {Patri, Adarsh S. and Sakai, Akito and Lee, SungBin and  Paramekanti, Arun and Nakatsuji, Satoru and Kim, Yong Baek},
    title = {Unveiling hidden multipolar orders with magnetostriction},
    journal = {Nature Communications},
    volume = {10},
    year = {2019},
    doi = {10.1038/s41467-019-11913-3}
}

@article{Tang2024,
  author  = {Tang, Nan and Gen, Masaki and Rotter, Martin and Man, Huiyuan and Matsuhira, Kazuyuki and Matsuo, Akira and Kindo, Koichi and Ikeda, Akihiko and Matsuda, Yasuhiro H. and Gegenwart, Philipp and others},
  title   = {Crystal field magnetostriction of spin ice under ultrahigh magnetic fields},
  journal = {Phys. Rev. B},
  volume  = {110},
  pages   = {214414},
  year    = {2024},
  doi     = {10.1103/PhysRevB.110.214414}
}

@book{Dresselhaus2008,
  author    = {Dresselhaus, Mildred S. and Dresselhaus, Gene and Jorio, Ado},
  title     = {Group Theory: Application to the Physics of Condensed Matter},
  publisher = {Springer},
  year      = {2008},
  address   = {Berlin, Heidelberg},
  isbn      = {978-3-540-32899-5},
  doi       = {10.1007/978-3-540-32899-5}
}

@article{OnodaTanaka2010,
  author  = {Onoda, Shigeki and Tanaka, Yoichi},
  title   = {Quantum Melting of Spin Ice: Emergent Cooperative Quadrupole and Chirality},
  journal = {Phys. Rev. Lett.},
  volume  = {105},
  pages   = {047201},
  year    = {2010},
  doi     = {10.1103/PhysRevLett.105.047201}
}

@article{OnodaTanaka2011,
  author  = {Onoda, Shigeki and Tanaka, Yoichi},
  title   = {Quantum fluctuations in the effective pseudospin-{$\frac{1}{2}$} model for magnetic pyrochlore oxides},
  journal = {Phys. Rev. B},
  volume  = {83},
  pages   = {094411},
  year    = {2011},
  doi     = {10.1103/PhysRevB.83.094411}
}

@article{SavaryBalents2017,
  author  = {Savary, Lucile and Balents, Leon},
  title   = {Disorder-Induced Quantum Spin Liquid in Spin Ice Pyrochlores},
  journal = {Phys. Rev. Lett.},
  volume  = {118},
  pages   = {087203},
  year    = {2017},
  doi     = {10.1103/PhysRevLett.118.087203}
}

@article{Wen2017,
  author  = {Wen, J.-J. and Koohpayeh, S. M. and Ross, K. A. and Trump, B. A. and McQueen, T. M. and Kimura, K. and Nakatsuji, S. and Qiu, Y. and Pajerowski, D. M. and Copley, J. R. D. and others},
  title   = {Disordered Route to the {{Coulomb}} Quantum Spin Liquid: Random Transverse Fields on Spin Ice in {{Pr$_2$Zr$_2$O$_7$}}},
  journal = {Phys. Rev. Lett.},
  volume  = {118},
  pages   = {107206},
  year    = {2017},
  doi     = {10.1103/PhysRevLett.118.107206}
}

@article{Benton2018,
  author  = {Benton, Owen},
  title   = {Instabilities of a {{U(1)}} Quantum Spin Liquid in Disordered Non-{{Kramers}} Pyrochlores},
  journal = {Phys. Rev. Lett.},
  volume  = {121},
  pages   = {037203},
  year    = {2018},
  doi     = {10.1103/PhysRevLett.121.037203}
}

@article{Onimaru2016,
  author  = {Onimaru, Takahiro and Kusunose, Hiroaki},
  title   = {Exotic Quadrupolar Phenomena in Non-{{Kramers}} Doublet Systems --- {{The}} Cases of {{Pr$T_2$Zn$_{20}$}} ({$T$} = {{Ir}}, {{Rh}}) and {{Pr$T_2$Al$_{20}$}} ({$T$} = {{V}}, {{Ti}}) ---},
  journal = {J. Phys. Soc. Jpn.},
  volume  = {85},
  pages   = {082002},
  year    = {2016},
  doi     = {10.7566/JPSJ.85.082002}
}

@article{Tang2023,
  author  = {Tang, Nan and Gritsenko, Yulia and Kimura, Kenta and Bhattacharjee, Subhro and Sakai, Akito and Fu, Mingxuan and Takeda, Hikaru and Man, Huiyuan and Sugawara, Kento and Matsumoto, Yosuke and others},
  title   = {Spin--orbital liquid state and liquid--gas metamagnetic transition on a pyrochlore lattice},
  journal = {Nat. Phys.},
  volume  = {19},
  pages   = {92},
  year    = {2023},
  doi     = {10.1038/s41567-022-01816-4}
}

@article{Bleaney1973,
  author  = {Bleaney, B.},
  title   = {Enhanced nuclear magnetism},
  journal = {Physica},
  volume  = {69},
  pages   = {317},
  year    = {1973},
  doi     = {10.1016/0031-8914(73)90224-3}
}

@article{Blote1969,
  author  = {Bl{\"o}te, H. W. J. and Wielinga, R. F. and Huiskamp, W. J.},
  title   = {Heat-capacity measurements on rare-earth double oxides {{R$_2$M$_2$O$_7$}}},
  journal = {Physica},
  volume  = {43},
  pages   = {549},
  year    = {1969},
  doi     = {10.1016/0031-8914(69)90187-6}
}

@article{Gronemann2023,
  title = {Impact of hyperfine contributions on the ground state of spin-ice compounds},
  author = {Gronemann, J. and Chattopadhyay, S. and Gottschall, T. and Osmic, E. and Islam, A. T. M. N. and Anand, V. K. and Lake, B. and Kaneko, H. and Suzuki, H. and Wosnitza, J. and Herrmannsd\"orfer, T.},
  journal = {Phys. Rev. B},
  volume = {108},
  issue = {21},
  pages = {214412},
  numpages = {8},
  year = {2023},
  month = {Dec},
  publisher = {American Physical Society},
  doi = {10.1103/PhysRevB.108.214412},
  url = {https://link.aps.org/doi/10.1103/PhysRevB.108.214412}
}

@article{Canals1998,
  author  = {Canals, B. and Lacroix, C.},
  title   = {Pyrochlore Antiferromagnet: A Three-Dimensional Quantum Spin Liquid},
  journal = {Phys. Rev. Lett.},
  volume  = {80},
  pages   = {2933},
  year    = {1998},
  doi     = {10.1103/PhysRevLett.80.2933}
}

@article{Udagawa2019,
  author  = {Udagawa, Masafumi and Moessner, Roderich},
  title   = {Spectrum of Itinerant Fractional Excitations in Quantum Spin Ice},
  journal = {Phys. Rev. Lett.},
  volume  = {122},
  pages   = {117201},
  year    = {2019},
  doi     = {10.1103/PhysRevLett.122.117201}
}

@article{Wei2023,
  author  = {Wei, C. and Curnoe, S. H.},
  title   = {Exact diagonalization for a 16-site spin-1/2 pyrochlore cluster},
  journal = {J. Phys.: Condens. Matter},
  volume  = {35},
  pages   = {295802},
  year    = {2023},
  doi     = {10.1088/1361-648X/acccc8}
}

@article{Wei2024,
  author  = {Wei, C. and Curnoe, S. H.},
  title   = {Symmetry considerations in exact diagonalization: spin-1/2 pyrochlore magnets},
  journal = {J. Phys. A: Math. Theor.},
  volume  = {57},
  pages   = {435301},
  year    = {2024},
  doi     = {10.1088/1751-8121/ad7fa7}
}

@article{Applegate2012,
  author  = {Applegate, R. and Hayre, N. R. and Singh, R. R. P. and Lin, T. and Day, A. G. R. and Gingras, M. J. P.},
  title   = {Vindication of {{Yb$_2$Ti$_2$O$_7$}} as a Model Exchange Quantum Spin Ice},
  journal = {Phys. Rev. Lett.},
  volume  = {109},
  pages   = {097205},
  year    = {2012},
  doi     = {10.1103/PhysRevLett.109.097205}
}

@article{Hayre2013,
  author  = {Hayre, N. R. and Ross, K. A. and Applegate, R. and Lin, T. and Singh, R. R. P. and Gaulin, B. D. and Gingras, M. J. P.},
  title   = {Thermodynamic properties of {{Yb$_2$Ti$_2$O$_7$}} pyrochlore as a function of temperature and magnetic field: Validation of a quantum spin ice exchange {{Hamiltonian}}},
  journal = {Phys. Rev. B},
  volume  = {87},
  pages   = {184423},
  year    = {2013},
  doi     = {10.1103/PhysRevB.87.184423}
}

@article{Schafer2020,
  author  = {Sch{\"a}fer, Robin and Hagym{\'a}si, Imre and Moessner, Roderich and Luitz, David J.},
  title   = {Pyrochlore {$S=\frac{1}{2}$} {{Heisenberg}} antiferromagnet at finite temperature},
  journal = {Phys. Rev. B},
  volume  = {102},
  pages   = {054408},
  year    = {2020},
  doi     = {10.1103/PhysRevB.102.054408}
}

@article{Schafer2023,
  author  = {Sch{\"a}fer, Robin and Placke, Benedikt and Benton, Owen and Moessner, Roderich},
  title   = {Abundance of Hard-Hexagon Crystals in the Quantum Pyrochlore Antiferromagnet},
  journal = {Phys. Rev. Lett.},
  volume  = {131},
  pages   = {096702},
  year    = {2023},
  doi     = {10.1103/PhysRevLett.131.096702}
}

@misc{SM,
  note = {See the Supplemental Material appended below for
          the weak-field derivation of $B_\mu(\bm p)$ and the acoustic/optical
          sublattice kernel, the exact band extraction and zero-transport
          projection, the photon- and spinon-window measurements and the
          acoustic-cancellation mechanism, the power-normalized
          $E_g$/$T_{2g}$/field comparison, and the strain-tensor irrep
          decomposition}
}

@article{Zhou2024DOfield,
  author  = {Zhou, Zhengbang and Desrochers, F\'elix and Kim, Yong Baek},
  title   = {Magnetic field response of dipolar-octupolar quantum spin ice},
  journal = {Phys. Rev. B},
  volume  = {110},
  pages   = {174441},
  year    = {2024},
  doi     = {10.1103/PhysRevB.110.174441}
}

@article{Zhou2025PhaseDiagram,
  author  = {Zhou, Zhengbang and Kim, Yong Baek},
  title   = {Towards a global phase diagram of {Ce}-based dipolar-octupolar
             pyrochlore magnets under magnetic fields},
  journal = {Phys. Rev. B},
  volume  = {112},
  pages   = {L060407},
  year    = {2025},
  doi     = {10.1103/8mhy-qwsw}
}

@article{Gao2026Demarcation,
  author  = {Gao, Bin and Zhou, Zhengbang and Zhang, Tingjun and
             Podlesnyak, Andrey and Cheong, Sang-Wook and Kim, Yong Baek and
             Dai, Pengcheng},
  title   = {Spectroscopic Demarcation of Emergent Photons and Spinons in a
             Dipolar-Octupolar Quantum Spin Liquid},
  journal = {Phys. Rev. Lett.},
  volume  = {136},
  pages   = {256703},
  year    = {2026},
  doi     = {10.1103/svt2-m3pp}
}

@article{Zhou2025QFI,
  author  = {Zhou, Chengkang and Zhou, Zhengbang and Desrochers, F{\'e}lix and Kim, Yong Baek and Meng, Zi Yang},
  title   = {Quantum {Fisher} Information as a Thermal Probe in Frustrated Magnets through Insights from Quantum Spin Ice},
  journal = {Nat. Commun.},
  year    = {2026},
  doi     = {10.1038/s41467-026-74589-6}
}

@article{Gao2025Photons,
  author  = {Gao, Bin and Desrochers, F{\'e}lix and Tam, David W. and Kirschbaum, Diana M. and Steffens, Paul and Hiess, Arno and Nguyen, Duy Ha and Su, Yixi and Cheong, Sang-Wook and Paschen, Silke and Kim, Yong Baek and Dai, Pengcheng},
  title   = {Neutron scattering and thermodynamic evidence for emergent photons and fractionalization in a pyrochlore spin ice},
  journal = {Nat. Phys.},
  volume  = {21},
  pages   = {1203--1210},
  year    = {2025},
  doi     = {10.1038/s41567-025-02922-9}
}

@article{Marinho2026,
  author  = {Marinho, Marcus V. and Andrade, Eric C.},
  title   = {Quantum spin liquids stabilized by disorder in non-{{Kramers}} pyrochlores},
  journal = {Ann. Phys. (Berlin)},
  volume  = {538},
  pages   = {e00552},
  year    = {2026},
  doi     = {10.1002/andp.202500552}
}

@article{Simon2022,
  author  = {Simon, Sophia and Patri, Adarsh S. and Kim, Yong Baek},
  title   = {Ultrasound detection of emergent photons in generic quantum spin ice},
  journal = {Phys. Rev. B},
  volume  = {106},
  pages   = {064427},
  year    = {2022},
  doi     = {10.1103/PhysRevB.106.064427}
}

@article{Desrochers2024FiniteT,
  author  = {Desrochers, F{\'e}lix and Kim, Yong Baek},
  title   = {Finite-temperature dynamics in {$0$}-flux and {$\pi$}-flux quantum spin
             ice: Self-consistent exclusive boson approach},
  journal = {Phys. Rev. B},
  volume  = {109},
  pages   = {144410},
  year    = {2024},
  doi     = {10.1103/PhysRevB.109.144410}
}

@article{Chern2024,
  author  = {Chern, Li Ern and Desrochers, F{\'e}lix and Kim, Yong Baek and
             Castelnovo, Claudio},
  title   = {Pseudofermion functional renormalization group study of
             dipolar-octupolar pyrochlore magnets},
  journal = {Phys. Rev. B},
  volume  = {109},
  pages   = {184421},
  year    = {2024},
  doi     = {10.1103/PhysRevB.109.184421}
}

@article{Sibille2020,
  author  = {Sibille, Romain and Gauthier, Nicolas and Lhotel, Elsa and
             Por{\'e}e, Victor and Pomjakushin, Vladimir and Ewings, Russell A. and
             Perring, Toby G. and Ollivier, Jacques and Wildes, Andrew and
             Ritter, Clemens and Hansen, Thomas C. and Keen, David A. and
             Nilsen, G{\o}ran J. and Keller, Lukas and Petit, Sylvain and
             Fennell, Tom},
  title   = {A quantum liquid of magnetic octupoles on the pyrochlore lattice},
  journal = {Nat. Phys.},
  volume  = {16},
  pages   = {546},
  year    = {2020},
  doi     = {10.1038/s41567-020-0827-7}
}

@article{Pace2021,
  author  = {Pace, Salvatore D. and Morampudi, Siddhardh C. and
             Moessner, Roderich and Laumann, Chris R.},
  title   = {Emergent Fine Structure Constant of Quantum Spin Ice Is Large},
  journal = {Phys. Rev. Lett.},
  volume  = {127},
  pages   = {117205},
  year    = {2021},
  doi     = {10.1103/PhysRevLett.127.117205}
}

@article{Seth2022,
  author  = {Seth, Arnab and Bhattacharjee, Subhro and Moessner, Roderich},
  title   = {Probing emergent {QED} in quantum spin ice via {Raman} scattering
             of phonons: Shallow inelastic scattering and pair production},
  journal = {Phys. Rev. B},
  volume  = {106},
  pages   = {054507},
  year    = {2022},
  doi     = {10.1103/PhysRevB.106.054507}
}

@article{Smith2025Octupolar,
  author  = {Smith, E. M. and Sch{\"a}fer, R. and Dudemaine, J. and
             Placke, B. and Yuan, B. and Morgan, Z. and Ye, F. and
             Moessner, R. and Benton, O. and Bianchi, A. D. and Gaulin, B. D.},
  title   = {Single-Crystal Diffuse Neutron Scattering Study of the
             Dipole-Octupole Quantum Spin-Ice Candidate {{Ce$_2$Zr$_2$O$_7$}}:
             No Apparent Octupolar Correlations Above {$T=0.05$~K}},
  journal = {Phys. Rev. X},
  volume  = {15},
  pages   = {021033},
  year    = {2025},
  doi     = {10.1103/PhysRevX.15.021033}
}

@article{Yuan2026CeSn,
  author  = {Yuan, Bo and Powell, M. and Liu, X. and Ni, J. and Smith, E. M.
             and Sch{\"a}fer, R. and Moessner, R. and Ye, F. and
             Dudemaine, J. and Bianchi, A. D. and Kolis, J. W. and
             Gaulin, B. D.},
  title   = {Observation of dipolar spin-ice-like correlations in the quantum
             spin ice candidate {{Ce$_2$Sn$_2$O$_7$}}},
  journal = {arXiv:2601.20766},
  year    = {2026},
  doi     = {10.48550/arXiv.2601.20766}
}

@article{Luo2025Pr,
  author  = {Luo, Yi and Ortiz, Brenden R. and Knudtson, Miles and
             Wilson, Stephen D. and Liu, Jue and Frandsen, Benjamin A. and
             Chen, Si Athena and Frontzek, Matthias D. and
             Podlesnyak, Andrey A. and Paddison, Joseph A. M. and
             Aczel, Adam A.},
  title   = {Disorder-induced proximate quantum spin ice phase in
             {{Pr$_2$Sn$_2$O$_7$}}},
  journal = {arXiv:2508.19248},
  year    = {2025},
  doi     = {10.48550/arXiv.2508.19248}
}

@article{Chung2025,
  author  = {Chung, Kristian Tyn Kai and Petit, Sylvain and Robert, Julien and
             McClarty, Paul},
  title   = {Geometrically frustrated quadrupoles on the pyrochlore lattice and
             generalized spin liquids},
  journal = {arXiv:2506.19908},
  year    = {2025},
  doi     = {10.48550/arXiv.2506.19908}
}

@article{Schaden2024,
  author  = {Schaden, Yannik and Gonzalez, Mat{\'\i}as G. and Reuther, Johannes},
  title   = {Phase diagram of the {{XXZ}} pyrochlore model from pseudo-{{Majorana}}
             functional renormalization group},
  journal = {Phys. Rev. B},
  volume  = {111},
  pages   = {134442},
  year    = {2025},
  doi     = {10.1103/PhysRevB.111.134442}
}
\SMtocon

\clearpage
\onecolumngrid

\setcounter{secnumdepth}{3}
\setcounter{tocdepth}{2}
\setcounter{section}{0}
\setcounter{equation}{0}
\setcounter{figure}{0}
\setcounter{table}{0}
\renewcommand{\thefigure}{S\arabic{figure}}
\renewcommand{\thetable}{S\Roman{table}}
\renewcommand{\theequation}{S\arabic{equation}}
\renewcommand{\thesection}{S\arabic{section}}
\renewcommand{\thesubsection}{\thesection.\Alph{subsection}}
\makeatletter\renewcommand{\p@subsection}{}\renewcommand{\p@subsubsection}{}\makeatother

\begin{center}
{\large\bfseries Supplemental Material for\\[2pt]
``Thermodynamic Spectroscopy of Emergent Excitations in Quantum Spin Ice''}\\[10pt]
Zhengbang Zhou, Tony An, and Yong Baek Kim\\[2pt]
\emph{Department of Physics, University of Toronto, Toronto, Ontario M5S 1A7, Canada}
\end{center}

\tableofcontents

\newpage


\section{Model, notation, and the two sectors}
\label{sec:model}

\subsection{Hamiltonian and lattice}

The pyrochlore lattice is a network of corner-sharing tetrahedra and decomposes into four interpenetrating face-centered-cubic sublattices. Throughout this Supplemental Material, we use Greek letters $\mu,\nu,\rho,\sigma\in\{0,1,2,3\}$ exclusively as sublattice indices, while Latin letters $i,j,\ldots$ label individual lattice sites. We denote by $\mu(i)$ the sublattice containing site $i$. Every tetrahedron contains one site from each sublattice. At site $i$ we use an effective spin-$1/2$ operator $\bm S_i$ expressed in its local frame, with the local $z$ axis chosen along the sublattice-dependent Ising direction $\hat{\bm n}_{\mu(i)}\parallel\langle111\rangle$. The four local axes obey
\begin{equation}
  \sum_{\mu=0}^{3}\hat{\bm n}_\mu=\bm 0.
  \label{eq:axessum}
\end{equation}
The model is the nearest-neighbor XXZ Hamiltonian of the main text,
\begin{equation}
  H_{\rm XXZ}=\Jzz\sum_{\langle ij\rangle}S^z_iS^z_j
  -\Jpm\sum_{\langle ij\rangle}\big(S^+_iS^-_j+S^-_iS^+_j\big),
  \label{eq:HxxzSM}
\end{equation}
We take $\Jzz>0$ and restrict to $|\Jpm|\leq0.3\Jzz$; all numerical results use $\Jzz=1$. Equation~\eqref{eq:HxxzSM} is the minimal nearest-neighbor model widely used for pyrochlore quantum-spin-ice candidates, from early analyses of $\mathrm{Yb_2Ti_2O_7}$~\cite{Ross2011,Applegate2012,Hayre2013} to the Pr- and Ce-based materials discussed in the main text. Its Coulomb phase has been established using quantum Monte Carlo~\cite{Banerjee2008,Shannon2012,Kato2015,Huang2018,Zhou2025QFI}, functional renormalization group~\cite{Chern2024,Schaden2024}, and lattice gauge theory~\cite{Hermele2004,Savary2012,Benton2012,Savary2013,Desrochers2023}.

\subsection{Charge, ice manifold, and the two sectors}

For each tetrahedron $t$ we define the charge $Q_t=\sum_{i\in t}S_i^z$. The Ising term in~\eqref{eq:HxxzSM} can be written 
\begin{equation}
  H_{\Jzz}=\frac{\Jzz}{2}\sum_t Q_t^2 
  \label{eq:chargehamPT}
\end{equation}
up to an additive constant which we drop. Therefore, states satisfying $Q_t=0$ on every tetrahedron form the ice manifold, and $\Pice$ denotes the corresponding projector. Configurations with $Q_t\neq0$ contain charged spinon defects and cost an energy of order $\Jzz$. This is the quantum analogue of the classical spin-ice Coulomb phase and its deconfined monopole excitations~\cite{Castelnovo2008,Jaubert2009,Fennell2009,Morris2009,Petit2016,Castelnovo2012}. We refer to the low-energy band adiabatically descended from the ice manifold as the \emph{ring-exchange sector}, and to its complement as the \emph{spinon sector}. The two thermal regimes emphasized below are the ring-exchange window, $T\sim |g|$, and the spinon window, $T\sim\Jzz/2$, where
\begin{equation}
  g=\frac{12\Jpm^3}{\Jzz^2}
  \label{eq:gSM}
\end{equation}
is the signed hexagonal ring-exchange coupling derived in Sec.~\ref{sec:pt:ring}. Its sign distinguishes the two flux sectors: $g>0$ for $\Jpm>0$ and $g<0$ for $\Jpm<0$.

\subsection{Field and probe patterns}

The transverse field of the main text is
\begin{equation}
  H(\lambda)=H_{\rm XXZ}-\lambda\Xhat,
  \qquad
  \Xhat=\sum_i\big(p_{\mu(i)}S^+_i+p_{\mu(i)}^{*}S^-_i\big),
  \label{eq:Xsub}
\end{equation}
where $\lambda$ is an energy-like generalized field and $p_{\mu(i)}$ is constant within each sublattice. We reserve subscripts on $p$ for sublattice labels and superscripts for probe channels. The probe is completely specified by the four-component pattern $\bm p=(p_0,p_1,p_2,p_3)$. For all probe patterns used below, a single global $S^z$ rotation makes the four amplitudes real; in that representative one may write $\Xhat=\sum_\mu p_\mu\Xhat_\mu$ with $\Xhat_\mu=\sum_{i\in\mu}(S_i^++S_i^-)$. An overall complex phase is therefore physically redundant for the observables considered here. The three patterns used below are
\begin{equation}
  \bm p^{E_g}=(1,1,1,1),
  \qquad
  \bm p^{T_{2g}}=\tfrac14(1,-1,1,-1),
  \label{eq:strainvectorsSM}
\end{equation}
and the magnetic field pattern, obtained from $p^{\hat{\bm B}}_\mu=\hat{\bm n}_\mu\!\cdot\!\hat{\bm B}$,
\begin{equation}
  \bm p^{[111]}=\big(1,-\tfrac13,-\tfrac13,-\tfrac13\big),
  \label{eq:fieldvectorsSM}
\end{equation}
A uniform laboratory magnetic field therefore produces four generally different sublattice amplitudes because the local Ising axes are different. By cubic symmetry $[001]$ is equivalent to $[100]$. Equation~\eqref{eq:strainvectorsSM} displays one member of the $T_{2g}$ triplet; the remaining two follow by point-group operations. 

\subsection{Response and normalization}

The central observable is the temperature derivative of the generalized polarization conjugate to $\lambda$,
\begin{equation}
  \gamma_X=\frac{d\expv{\Xhat}}{dT}=\frac{1}{T^2}\Cov_T(\Xhat,H),
  \qquad
  \Cov_T(A,B)=\expv{AB}_T-\expv{A}_T\expv{B}_T,
  \label{eq:maxwellSM}
\end{equation}
with all thermal averages evaluated using $H(\lambda)$. Depending on the physical realization, $\gamma_X$ becomes the symmetry-resolved thermal-expansion response or the magnetocaloric derivative $dM/dT$ of the main text. We use the common symbol $\gamma_X$ whenever the derivation applies to both. To compare probe patterns with different overall amplitudes, we normalize by the form factor magnitutde
\begin{equation}
  \Ppow=\sum_i|p_i|^2=\frac{N}{4}\sum_\mu|p_\mu|^2 ,
  \label{eq:Rnorm}
\end{equation}
with $N=16$ the number of sites; $\Ppow=16$, $1$ and $16/3$ for $\bm p^{E_g}$, $\bm p^{T_{2g}}$ and $\bm p^{[111]}$ respectively. Field strengths are quoted through the dimensionless combination
\begin{equation}
  x\equiv\frac{\lambda^2}{|\Jpm|\Jzz},
  \label{eq:xdef}
\end{equation}
which is the natural expansion parameter of Eq.~\eqref{eq:geffSM} below. Unless another value is stated, numbers refer to $\Jpm=-0.10$ ($\pi$ flux) at $x=0.014$ ($\lambda=0.0374$).

\section{Weak-field theory of the ring-exchange sector}
\label{sec:pt}

\subsection{The ring exchange and its energy scale}
\label{sec:pt:ring}
Our approach is to construct an effective Hamiltonian that acts only within the ice manifold, following the treatment given in ~\cite{Hermele2004,Sanders2024,SandersYan2024}. We use the perturbation expansion
\begin{equation}
    H_{\rm eff} = \Pice V \sum_{k \geq 0} \left(\frac{1-\Pice}{-H_0} V\right)^k \Pice \label{eq:Heff_expansion}
\end{equation}
where 
\begin{align}
    H_0 &= \frac{\Jzz}{2} \sum_t Q_t^2 \\
    V &= -\Jpm\sum_{\langle ij\rangle}\big(S^+_iS^-_j+S^-_iS^+_j\big). 
\end{align}
Each term in Eq.~\eqref{eq:Heff_expansion} can be interpreted as a process which starts in the ice manifold (ground state), excites virtual spinons through the transverse exchange $V$, and then annihilates them to return to the ice manifold. This requires a sequence of exchanges whose net set of flipped spins forms a closed alternating loop. The contribution from each process is weighted by energy denominators $1/H_0$ which account for the energy of every intermediate state in the sequence.

On the infinite pyrochlore lattice the shortest such loop is a hexagon. There are four hexagon orientations, labeled by $\mu\in\{0,1,2,3\}$. A hexagon of orientation $\mu$ omits sublattice $\mu$ and visits the remaining three sublattices, denoted $\nu,\rho,\sigma$, in the cyclic order
\begin{equation}
  \nu,\rho,\sigma,\nu,\rho,\sigma,
  \qquad
  \{\nu,\rho,\sigma\}=\{0,1,2,3\}\setminus\{\mu\}.
\end{equation}
For a representative flippable hexagon of orientation $\mu$, label the six sites cyclically so that the directed plaquette flip is
\begin{equation}
  W_{\mu,\hexagon} = S^+_1S^-_2S^+_3S^-_4S^+_5S^-_6 .
  \label{eq:Whex}
\end{equation}
We then define
\begin{equation}
  W_\mu = \sum_{\hexagon\in\mu}W_{\mu,\hexagon},
  \label{eq:Wmu}
\end{equation}
where the sum runs over all hexagons of orientation $\mu$. Thus $W_\mu+W_\mu^\dagger$ is the Hermitian ring-exchange operator for that orientation.

The numerical factor in the bare ring amplitude follows directly from the third-order term in~\eqref{eq:Heff_expansion}. The six sites of a flippable hexagon can be covered by exchange bonds in two perfect matchings,
\begin{equation}
  (12)(34)(56),
  \qquad
  (23)(45)(61).
  \label{eq:hexmatchings}
\end{equation}
For either matching, the three exchange operators can occur in $3!=6$ time orderings. After the first and second exchanges the state contains one spinon pair, so both intermediate states lie an energy $\Jzz$ above the ice manifold. Including the sign of the three exchange vertices gives the effective plaquette Hamiltonian
\begin{equation}
  H_{\rm ring}
  =
  -g\sum_{\mu=0}^{3}
  \left(W_\mu+W_\mu^\dagger\right),
  \qquad
  g=\frac{12\Jpm^3}{\Jzz^2}.
  \label{eq:ringSM}
\end{equation}
There is therefore only one ring coupling, and it is signed. For $\Jpm>0$, $g>0$ and the ring term favors the $0$-flux sector; for $\Jpm<0$, $g<0$ and it favors the $\pi$-flux sector. The characteristic low-energy scale is $|g|$, while the sign of $g$ carries the flux information. Coarse-grained, Eq.~\eqref{eq:ringSM} gives compact $U(1)$ lattice electrodynamics with a gapless emergent photon whose bandwidth is of order $|g|$~\cite{Hermele2004,Savary2012,Lee2012,Benton2018Pi}.

\subsection{The field correction to the ring exchange}
\label{sec:pt:B}

In this section we carry out the expansion~\eqref{eq:Heff_expansion} including the transverse field $-\lambda\Xhat$ in the perturbation $V$. A transverse field has no matrix element within the ice manifold,
\begin{equation}
  \Pice\Xhat\Pice=0 ,
  \label{eq:P0XP0SM}
\end{equation}
because a single $S_i^\pm$ changes the charge on both tetrahedra sharing site $i$. The leading field dependence of the ice-sector Hamiltonian is therefore even in $\lambda$ and begins at order $\lambda^2$.

The first field-induced process that produces the same six-spin plaquette flip as $W_{\mu,\hexagon}$ contains two field insertions and two transverse-exchange insertions. It scales as
\begin{equation}
  \frac{\lambda^2\Jpm^2}{\Jzz^3}.
\end{equation}
Relative to the bare coupling $g\sim\Jpm^3/\Jzz^2$, this gives the dimensionless ratio $\lambda^2/(\Jpm\Jzz)$. The sign of this ratio is already the sign information we need: no separate unsigned ring scale is required.

For a partially flipped alternating loop, each connected flipped segment has two endpoints and therefore carries one spinon pair. A virtual configuration containing $n_{\rm seg}$ disconnected flipped segments consequently lies
\begin{equation}
  \Delta E=n_{\rm seg}\Jzz
  \label{eq:segmentenergy}
\end{equation}
above the ice manifold by Eq.~\eqref{eq:chargehamPT}. This rule fixes all energy denominators below.

Each of the six sites of the final plaquette flip must be flipped exactly once. A transverse exchange flips two neighboring sites, whereas a field insertion flips one site. Two field vertices and two exchange vertices must therefore tile the six-cycle by two single-site tiles and two disjoint nearest-neighbor edges. If the two field insertions act on sites $i$ and $j$, the remaining four sites can be covered by two disjoint exchange edges if and only if $i$ and $j$ are adjacent or antipodal:
\begin{equation}
  \{i,j\}\in
  \underbrace{\{12,23,34,45,56,61\}}_{\text{6 adjacent pairs}}
  \ \cup\
  \underbrace{\{14,25,36\}}_{\text{3 antipodal pairs}} .
  \label{eq:tilings}
\end{equation}
Thus there are nine allowed tilings.

For the directed plaquette operator in Eq.~\eqref{eq:Whex}, odd sites are raised and even sites are lowered. Every pair in Eq.~\eqref{eq:tilings} contains one odd and one even site, so all nine tilings are compatible with the alternating spin structure. If $i$ is the odd field-flipped site and $j$ the even one, its form-factor weight is
\begin{equation}
  p_{\mu(i)}p_{\mu(j)}^* .
\end{equation}

The remaining numerical factor comes from summing the $4!$ time orderings of the four perturbing vertices. For any particular ordering, the four vertices generate three virtual intermediate states before the process returns to the ice manifold. We denote their excitation energies above the initial ice-manifold energy by $\Delta_1$, $\Delta_2$, and $\Delta_3$, where $\Delta_r$ is the energy of the virtual state after the first $r$ perturbing vertices have acted. These are therefore ordering-dependent intermediate-state energies, not three distinct fixed gaps of the Hamiltonian.

For a fixed tiling, the first vertex always produces one connected flipped segment and hence one spinon pair, so $\Delta_1=\Jzz$. After three vertices have acted, the complement is the single remaining tile; the flipped set is again connected, giving $\Delta_3=\Jzz$. Only the second intermediate state depends on the ordering. If the first two vertices form a connected flipped set, there is one spinon pair and $\Delta_2=\Jzz$; if they form two disconnected flipped segments, there are two spinon pairs and $\Delta_2=2\Jzz$.

Among the six unordered choices of the first two tiles, four give a connected union and two give two disconnected flipped segments. Each unordered pair corresponds to $2!\times2!=4$ complete time orderings, because the first two and last two vertices may each be interchanged. Therefore sixteen of the $24$ orderings have $(\Delta_1,\Delta_2,\Delta_3)=(\Jzz,\Jzz,\Jzz)$, while eight have $(\Delta_1,\Delta_2,\Delta_3)=(\Jzz,2\Jzz,\Jzz)$. For every allowed tiling,
\begin{equation}
\begin{split}
  \sum_{\rm orderings}
  \frac{1}{\Delta_1\Delta_2\Delta_3}
  &=
  \frac{16}{\Jzz^3}
  +\frac{8}{2\Jzz^3}=
  \frac{20}{\Jzz^3}.
  \label{eq:hexorderingfactor}
\end{split}
\end{equation}

Now consider a hexagon of orientation $\mu$. Its sublattice sequence is $\nu,\rho,\sigma,\nu,\rho,\sigma$, with $\{\nu,\rho,\sigma\}=\{0,1,2,3\}\setminus\{\mu\}$. Each of the three sublattices therefore occurs exactly once among the odd sites and once among the even sites. The sum over all nine allowed field pairs factorizes as
\begin{equation}
\begin{split}
  \sum_{\substack{i\ {\rm odd}\\ j\ {\rm even}}}
  p_{\mu(i)}p_{\mu(j)}^*
  &=
  (p_\nu+p_\rho+p_\sigma)
  (p_\nu+p_\rho+p_\sigma)^* 
  \label{eq:hexfactorization}
\end{split}
\end{equation}
The field therefore shifts the signed coupling of plaquettes of orientation $\mu$ by
\begin{equation}
  \delta g_\mu = 20 \left|\sum_{\nu\neq\mu}p_\nu\right|^2
  \frac{\lambda^2\Jpm^2}{\Jzz^3}.
  \label{eq:deltagmu}
\end{equation}
The orientation-dependent renormalized coupling is
\begin{equation}
  g_{\mu}^{\rm eff}(\lambda)
  =
  g+\delta g_\mu+O(\lambda^4),
  \label{eq:gmueffadditive}
\end{equation}
which may equivalently be written
\begin{equation}
  g_{\mu}^{\rm eff}(\lambda)
  =
  g\left[
    1+B_\mu(\bm p)\frac{\lambda^2}{\Jpm\Jzz}
    +O(\lambda^4)
  \right],
  \label{eq:gmueff}
\end{equation}
with
\begin{equation}
  B_\mu(\bm p)
  =
  \frac{5}{3}
  \left|\sum_{\nu\neq\mu}p_\nu\right|^2
  \geq0 .
  \label{eq:BmuSM}
\end{equation}
For the sublattice-uniform $E_g$ pattern, $p_\mu=1$, the three-sublattice sum equals $3$ for every orientation. Hence
\begin{equation}
  \delta g
  =
  20\times 3^2\,
  \frac{\lambda^2\Jpm^2}{\Jzz^3}
  =
  180\frac{\lambda^2\Jpm^2}{\Jzz^3},
  \label{eq:uniformdeltag}
\end{equation}
and all four orientations remain equivalent:
\begin{equation}
  H_{\rm ring}(\lambda)
  =
  -g^{\rm eff}(\lambda)
  \sum_{\mu=0}^{3}\left(W_\mu+W_\mu^\dagger\right),
  \label{eq:Hringeff}
\end{equation}
with the single signed dressed coupling
\begin{equation}
  \geff(\lambda)
  =
  g+\delta g
  =
  g\left[
    1+15\frac{\lambda^2}{\Jpm\Jzz}
    +O(\lambda^4)
  \right].
  \label{eq:geffSM}
\end{equation}
Thus
\begin{equation}
  B(\bm p^{E_g})=15.
\end{equation}

This signed form makes the physical effect immediate. Since $\delta g_\mu\geq0$, a field shifts every plaquette coupling toward more positive values. On the $0$-flux side, $g>0$, this increases $|g|$ and stiffens the ring sector. On the $\pi$-flux side, $g<0$, the same positive correction makes $g$ less negative, decreases $|g|$, and softens the ring sector. The flux dependence therefore resides directly in the sign of the single coupling $g$.

For a general probe the four orientations need not be renormalized equally. Cubic symmetry exchanges the four orientations, so the $O(\lambda^2)$ free energy samples their average
\begin{equation}
  \overline B(\bm p)=\frac14\sum_\mu B_\mu(\bm p).
\end{equation}
For uniform $E_g$ strain, however, every $B_\mu=15$, so the dressed ring Hamiltonian retains the one-coupling form in Eq.~\eqref{eq:Hringeff}.

\subsection{The response as a bilinear form}
\label{sec:pt:flavor}

The weak-field response can be organized independently of the loop expansion by resolving the probe in the four-dimensional sublattice space. This representation isolates all dependence on the form factor $\bm p$ and makes the relation between strain and magnetic probes transparent.

At $\lambda=0$ the response vanishes identically. The unitary operator
\begin{equation}
  G=\exp\left(i\pi\sum_iS_i^z\right)
\end{equation}
is a global $\pi$ rotation about the local $z$ axes. It leaves $H_{\rm XXZ}$ invariant but reverses every transverse spin component, so
\begin{equation}
  G\Xhat G^\dagger=-\Xhat .
\end{equation}
Field-free eigenstates can therefore be chosen with definite $G$ parity, implying $\bra{n}\Xhat\ket{n}=0$ for every eigenstate. Consequently $\expv{\Xhat}=0$ and $\Cov(\Xhat,H_{\rm XXZ})=0$ at $\lambda=0$. The same symmetry makes the free energy even in $\lambda$, and its weak-field expansion is
\begin{equation}
  F(\lambda,T)
  =
  F(0,T)
  -\frac{\lambda^2}{2}\bm p^\dagger\chi(T)\bm p
  +O(\lambda^4),
  \qquad
  \gamma_X
  =
  \lambda\,\bm p^\dagger\partial_T\chi(T)\bm p
  +O(\lambda^3).
  \label{eq:weakkernel}
\end{equation}
Here $\chi_{\mu\nu}(T)$ is the static susceptibility matrix of the four sublattice operators $\Xhat_\mu$,
\begin{equation}
  \chi_{\mu\nu}(T)
  =
  \int_0^\beta d\tau\,
  \expv{\delta\Xhat_\mu(\tau)\delta\Xhat_\nu(0)} .
\end{equation}
Microscopically, the leading response contains two insertions: one field insertion induces a transverse polarization and the measured operator supplies the second. The response is therefore bilinear in $\bm p$, with all pattern dependence carried by a single temperature-dependent $4\times4$ kernel.

Symmetry reduces this kernel to two scalar functions. Conservation of $S^z_{\rm tot}$ makes a global phase of $\bm p$ physically redundant, so the patterns considered here may be chosen real and $\chi$ may be taken real symmetric. Moreover, Eq.~\eqref{eq:HxxzSM} assigns identical couplings to every nearest-neighbor bond. Any bond-graph automorphism that permutes the four sublattices therefore leaves the Hamiltonian invariant and permutes the operators $\Xhat_\mu$ in the same way. On the $16$-site cluster these automorphisms induce all $24$ permutations of $\{0,1,2,3\}$. Because this action is $2$-transitive, an invariant matrix can distinguish only diagonal from off-diagonal entries:
\begin{equation}
  k_{\mu\nu}(T)
  \equiv
  \frac{4}{N}\partial_T\chi_{\mu\nu}(T)
  =
  \chi_d(T)\delta_{\mu\nu}
  +
  \chi_o(T)\big(1-\delta_{\mu\nu}\big).
  \label{eq:flavorkernel}
\end{equation}
Thus $\chi_d$ is the common same-sublattice response and $\chi_o$ is the common cross-sublattice response.

Using the normalization in Eq.~\eqref{eq:Rnorm},
\begin{equation}
  \frac{\gamma_X}{\lambda\Ppow}
  =
  \frac{\bm p^\dagger k\bm p}{\sum_\mu|p_\mu|^2}
  +O(\lambda^2)
  =
  \begin{cases}
    \chi_d+3\chi_o, & \bm p\ \text{sublattice uniform},\\[2pt]
    \chi_d-\chi_o, & \sum_\mu p_\mu=0.
  \end{cases}
  \label{eq:Reig}
\end{equation}
The uniform and zero-sum subspaces are therefore the two eigenspaces of the response kernel. In particular, all zero-sum probes have the same response per $\sum_i|p_i|^2$. This includes the $T_{2g}$ strain channel as well as any magnetic fields. Conversely, Eq.~\eqref{eq:axessum} guarantees that a magnetic field can never access the uniform channel. Among the probes considered here, that channel is realized by the sublattice-uniform strain coupling, which is only possible via the $E_g$ strain.

Equations~\eqref{eq:flavorkernel} and \eqref{eq:Reig} use only $S^z_{\rm tot}$ conservation, the $G$ parity, and sublattice permutation symmetry. They therefore hold for every $\Jpm$ and at every temperature; the only weak-field approximation is the truncation of Eq.~\eqref{eq:weakkernel}. What changes between physical regimes is not the form of the kernel but the values and relative signs of $\chi_d$ and $\chi_o$.

\subsection{Proportionality to the specific heat and the peak shift}
\label{sec:pt:lockstep}

For a general probe, the four ring orientations acquire couplings
\begin{equation}
  g_\mu^{\rm eff}(\lambda)
  =g\left[1+B_\mu(\bm p)\frac{\lambda^2}{\Jpm\Jzz}+O(\lambda^4)\right],
  \label{eq:gmugeneralLock}
\end{equation}
and it is useful to define the Hermitian orientation-resolved ring operator
\begin{equation}
  \mathcal W_\mu\equiv W_\mu+W_\mu^\dagger,
  \qquad
  H_{\rm ring}(\lambda)=-\sum_\mu g_\mu^{\rm eff}(\lambda)\mathcal W_\mu.
  \label{eq:HringgeneralLock}
\end{equation}
The generalized force restricted to this sector is therefore
\begin{equation}
  \Xhat\big|_{\rm ring}
  =-\partial_\lambda H_{\rm ring}
  =\sum_\mu \partial_\lambda g_\mu^{\rm eff}\,\mathcal W_\mu.
  \label{eq:XgeneralLock}
\end{equation}
Using Eq.~\eqref{eq:maxwellSM},
\begin{equation}
  \gamma_X\big|_{\rm ring}
  =-\sum_{\mu\nu}
  (\partial_\lambda g_\mu^{\rm eff})g_\nu^{\rm eff}
  \frac{\Cov_T(\mathcal W_\mu,\mathcal W_\nu)}{T^2}.
  \label{eq:gammageneralLock}
\end{equation}
At $\lambda=0$ all $g_\mu^{\rm eff}=g$. Cubic symmetry permutes the four orientations, so the row sum
\begin{equation}
  R(T)\equiv \sum_\nu
  \frac{\Cov_{0,T}(\mathcal W_\mu,\mathcal W_\nu)}{T^2}
  \label{eq:rowsumLock}
\end{equation}
is independent of $\mu$. The zero-field ring-sector specific heat is consequently
\begin{equation}
  C_{\rm ring}(T)=4g^2R(T).
  \label{eq:Cringrowsum}
\end{equation}
Since
\begin{equation}
  \partial_\lambda g_\mu^{\rm eff}
  =2gB_\mu(\bm p)\frac{\lambda}{\Jpm\Jzz}+O(\lambda^3),
\end{equation}
Eq.~\eqref{eq:gammageneralLock} gives
\begin{equation}
  \gamma_X\big|_{\rm ring}
  =-\frac{2\overline B(\bm p)\lambda}{\Jpm\Jzz}
  C_{\rm ring}(T)+O(\lambda^3),
  \qquad
  \overline B(\bm p)=\frac14\sum_\mu B_\mu(\bm p).
  \label{eq:lockstepGeneralSM}
\end{equation}
Here the covariance matrix and $C_{\rm ring}$ may equally well be evaluated at the finite weak field: their $O(\lambda^2)$ corrections change Eq.~\eqref{eq:lockstepGeneralSM} only at $O(\lambda^3)$. Defining
\begin{equation}
  \overline g_{\rm eff}
  \equiv \frac14\sum_\mu g_\mu^{\rm eff}
  =g\left[1+\overline B(\bm p)\frac{\lambda^2}{\Jpm\Jzz}+O(\lambda^4)\right],
\end{equation}
we can therefore write the main-text result as
\begin{equation}
  \gamma_X\big|_{\rm ring}
  =\Gph(\lambda)C_{\rm ring}(T)+O(\lambda^3),
  \qquad
  \Gph(\lambda)=-\partial_\lambda\ln|\overline g_{\rm eff}|.
  \label{eq:lockstepSM}
\end{equation}
This derivation also shows explicitly why the response sign and peak-shift sign are opposite: for $\lambda>0$ and $\overline B\geq0$, $\Gph\propto-1/\Jpm$.

The same symmetry argument controls the peak shift for a general orientation-dependent probe. Write $g_\mu^{\rm eff}=g+\delta g_\mu$, with $\delta g_\mu=O(\lambda^2)$. Expanding the ring-sector free energy about the isotropic point gives
\begin{align}
  F_{\rm ring}(\{g_\mu^{\rm eff}\},T)
  &=F_{\rm ring}(g,T)
   +\sum_\mu\left.\frac{\partial F_{\rm ring}}{\partial g_\mu}\right|_{g_\nu=g}\delta g_\mu
   +O(\lambda^4)\\
  &=F_{\rm ring}(\overline g_{\rm eff},T)+O(\lambda^4),
  \label{eq:FaverageLock}
\end{align}
where the second line follows because cubic symmetry makes the four derivatives equal at the isotropic point. Thus, to $O(\lambda^2)$, the thermodynamics is the same as a uniform rescaling $g\rightarrow\overline g_{\rm eff}$, and
\begin{equation}
  T_{\rm pk}(\lambda)
  =T_{\rm pk}(0)
  \left[1+\kappa\lambda^2+O(\lambda^4)\right],
  \qquad
  \kappa=\frac{\overline B(\bm p)}{\Jpm\Jzz}.
  \label{eq:kappaSM}
\end{equation}
Hence the peak moves upward for $\Jpm>0$ and downward for $\Jpm<0$, while the signed thermal response has the opposite sign.

For a sublattice-uniform probe, and in particular the $E_g$ pattern, all four plaquette orientations acquire exactly the same dressed coupling: $B_\mu=B$. In this special case the ring Hamiltonian retains a one-coupling form,
\begin{equation}
  H_{\mathrm{ring}}(\lambda)
  =\geff(\lambda)\widetilde H,
  \qquad
  \widetilde H=-\sum_{\mu=0}^{3}\mathcal W_\mu,
  \label{eq:Hgscale}
\end{equation}
where $\widetilde H$ is independent of $\lambda$. The operator conjugate to $\lambda$ then obeys the exact identity within this effective Hamiltonian,
\begin{equation}
  \Xhat\big|_{\mathrm{ring}}
  =-\partial_\lambda H_{\mathrm{ring}}
  =-\partial_\lambda\ln|\geff|\,H_{\mathrm{ring}}.
  \label{eq:Xisenergy}
\end{equation}
Consequently,
\begin{equation}
  \Cov_T(\Xhat,H_{\mathrm{ring}})
  =-\partial_\lambda\ln|\geff|\,\Var_T(H_{\mathrm{ring}})
  =-\partial_\lambda\ln|\geff|\,T^2C_{\rm ring}(T),
  \label{eq:covcollapse}
\end{equation}
and therefore
\begin{equation}
  \gamma_X\big|_{\mathrm{ring}}
  =-\partial_\lambda\ln|\geff|\,C_{\rm ring}(T).
  \label{eq:lockstepExactSM}
\end{equation}
Thus the lockstep relation is exact within the one-coupling ring Hamiltonian for the uniform $E_g$ probe; only the microscopic weak-field expansion of $\geff$ is perturbative. For the thermodynamic-limit hexagon, $B(\bm p^{E_g})=15$ from Eq.~\eqref{eq:geffSM}. On the periodic $16$-site cluster, the same argument applies to the boundary-winding four-loop with $g\to g_4$ and $B\to B_4=6$, as derived below.

\section{Numerical protocol}
\label{sec:method}

We evaluate $H(\lambda)=H_{\rm XXZ}-\lambda\Xhat$ by exact diagonalization on the periodic $16$-site cubic pyrochlore cluster. The small cell permits access to the complete spectrum but also introduces a particularly important finite-size effect: a noncontractible four-site loop appears below the physical hexagon scale. Section~\ref{sec:pt:cluster} derives this effect explicitly.

\subsection{Cluster, basis, and observables}
\label{sec:method:cluster}

The periodic cluster contains $16$ sites, $8$ tetrahedra, and $48$ nearest-neighbor bonds. Its ice manifold has dimension $\operatorname{rank}\Pice=90$~\cite{Canals1998,Schafer2020,Schafer2023,Wei2023,Wei2024}. Because the one-body operator $\Xhat$ changes $S^z_{\rm tot}$ by $\pm1$, the field couples different magnetization sectors. We therefore diagonalize in the full $2^{16}$-dimensional spin basis rather than block-diagonalizing by $S^z_{\rm tot}$.

The thermodynamic observables used here can be reconstructed from the field-dependent spectrum alone:
\begin{align}
  C(T)&=\frac{\Var_T(H)}{T^2}, \\
  \expv{\Xhat}&=T\partial_\lambda\ln Z,\\
  \gamma_X&=\frac{d\expv{\Xhat}}{dT},
\end{align}
The second equality is equivalent to $\expv{\Xhat}=-\partial F/\partial\lambda$. Numerically, we obtain the $\lambda$ derivative by fitting $\ln Z(\lambda)$ to an even polynomial using the four nodes $x\in\{0,0.007,0.014,0.021\}$. Evenness in $\lambda$ follows from the parity symmetry established in Sec.~\ref{sec:pt:flavor}. 

\subsection{The perturbative limit of the periodic 16-site cluster}
\label{sec:pt:cluster}

Every ice-to-ice process generated by the transverse exchange flips a closed alternating loop. For a loop of length $2k$, the minimal process contains $k$ exchange vertices and $k-1$ virtual energy denominators of order $\Jzz$, so its characteristic amplitude scales as
\begin{equation}
  |g_{2k}|\sim \frac{|\Jpm|^k}{\Jzz^{k-1}} .
\end{equation}
On the infinite pyrochlore lattice the shortest allowed ice-to-ice loop is the six-site hexagon, $k=3$.

\begin{figure*}[t]
\includegraphics[width=0.98\textwidth]{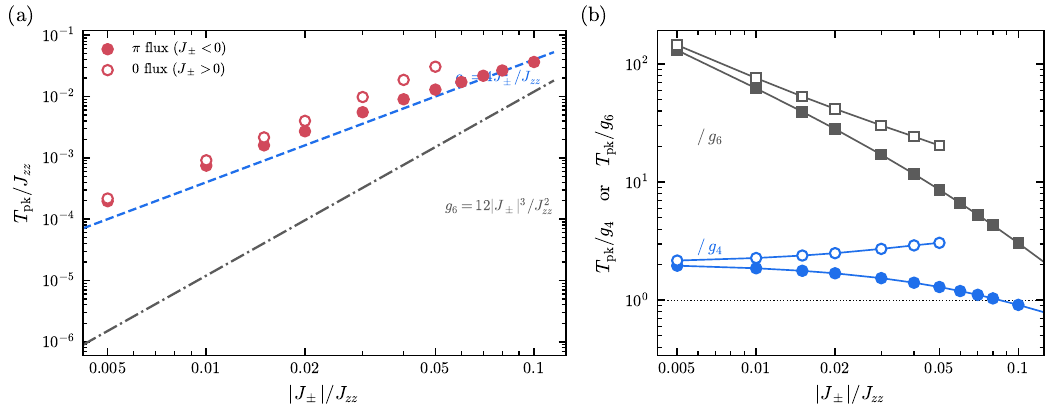}
\caption{The low-temperature scale of the periodic $16$-site cluster is set by a boundary-winding four-loop rather than the contractible hexagon. The low-temperature specific-heat peak $T_{\rm pk}$ is shown at $\lambda=0$; filled symbols denote $\pi$ flux ($\Jpm<0$) and open symbols $0$ flux ($\Jpm>0$). (a)~Comparison with the two candidate loop scales: the four-loop scale $g_4=4\Jpm^2/\Jzz$ (dashed) and the physical hexagon scale $|g|=12|\Jpm|^3/\Jzz^2$ (dot-dashed). The measured peak follows $g_4$. (b)~The same data divided by the two scales. As $|\Jpm|\to0$, $T_{\rm pk}/g_4$ approaches $1.96$ in the $\pi$-flux sector and $2.17$ in the $0$-flux sector, whereas $T_{\rm pk}/|g|$ diverges. The dotted line marks unity.}
\label{fig:peakscaling}
\end{figure*}

Periodic boundary conditions change this hierarchy on the $16$-site cubic cell. On a 16-site cubic cluster, a length-$4$ cycle can connect two two-in-two-out ice states. These cycles are noncontractible: they close only through the periodic identification and have no counterpart as local four-site ice-to-ice processes in the thermodynamic lattice.

The bare four-loop amplitude follows from second-order perturbation theory. Label the loop sites $1,\ldots,4$ cyclically. The four sites can be covered by exchange edges in two perfect matchings,
\begin{equation}
  (12)(34),
  \qquad
  (23)(41).
\end{equation}
For each matching, the two exchange operators can occur in $2!$ orders, and the single intermediate state contains one spinon pair with excitation energy $\Jzz$. Therefore
\begin{equation}
  g_4
  =
  2\times2!\,
  \frac{\Jpm^2}{\Jzz}
  =
  \frac{4\Jpm^2}{\Jzz}.
  \label{eq:g4}
\end{equation}

Figure~\ref{fig:peakscaling} confirms that this boundary-induced scale controls the low-temperature peak of the periodic cluster. The ratio between the lower peak position of the specific heat $T_{\pm pk}$ and $g_4$, $T_{\rm pk}/g_4$, tends to an $O(1)$ constant as $|\Jpm|\to0$, whereas $T_{\rm pk}/|g|$ diverges as $1/|\Jpm|$.

We next renormalize $g_4$ in the presence of $\lambda\Xhat$. The directed four-spin loop is
\begin{equation}
  W_4=S_1^+S_2^-S_3^+S_4^- .
\end{equation}
At leading order in the field, two sites are flipped by field vertices and the remaining two by one exchange vertex. The two field-flipped sites must consist of one odd and one even site,
\begin{equation}
  i\in\{1,3\},
  \qquad
  j\in\{2,4\},
\end{equation}
so there are four allowed field pairs. For every such choice, the two unflipped sites form the unique remaining edge of the loop.

A fixed tiling now contains three perturbing vertices: two one-site field operators and one exchange operator. They may occur in all $3!=6$ time orderings. After either the first or the second vertex has acted, the flipped sites form a single connected segment around the four-cycle. Both intermediate states therefore contain one spinon pair, and every ordering carries the same denominator $1/\Jzz^2$. The complete ordering sum is consequently
\begin{equation}
  \sum_{\rm orderings}
  \frac{1}{\Delta_1\Delta_2}
  =
  \frac{3!}{\Jzz^2}
  =
  \frac{6}{\Jzz^2}.
  \label{eq:fourorderingfactor}
\end{equation}

Summing over the four odd-even choices of the field vertices gives the factorized form
\begin{equation}
  \sum_{\substack{i=1,3\\j=2,4}}p_i p_j^*
  =
  (p_1+p_3)(p_2+p_4)^* .
  \label{eq:fourtilingsum}
\end{equation}
Thus the third-order correction to the four-loop Hamiltonian is
\begin{equation}
  \delta H_4^{(3)}
  =
  -3!\,
  \frac{\lambda^2\Jpm}{\Jzz^2}
  \left[
    (p_1+p_3)(p_2+p_4)^*W_4
    +\mathrm{h.c.}
  \right].
  \label{eq:4site_pt}
\end{equation}
The Hermitian-conjugate term contains the oppositely directed loop operator and the conjugated form factor.

For the uniform $E_g$ pattern, $p_i=1$, Eq.~\eqref{eq:4site_pt} gives 
\begin{equation}
  \delta g_4 = 3!\times4 \frac{\lambda^2\Jpm}{\Jzz^2} =
  24\frac{\lambda^2\Jpm}{\Jzz^2}.
\end{equation}
Relative to the bare factor $4$ in Eq.~\eqref{eq:g4},
\begin{equation}
  \frac{\delta g_4}{g_4} = 6\frac{\lambda^2}{\Jpm\Jzz}.
\end{equation}
The dressed four-loop coupling therefore takes the same form as the thermodynamic-limit ring coupling,
\begin{equation}
  g_4^{\rm eff}(\lambda)
  =
  g_4\left[
    1+B_4(\bm p)\frac{\lambda^2}{\Jpm\Jzz}
    +O(\lambda^4)
  \right],
  \label{eq:g4eff}
\end{equation}
with
\begin{equation}
  B_4(\bm p^{E_g})=6.
\end{equation}

The exact uniform-probe proportionality in Eq.~\eqref{eq:lockstepExactSM} therefore carries over unchanged after the replacements $g\to g_4$ and $B\to B_4$. Numerically, for the sublattice-uniform $E_g$ probe, $\kappa|\Jpm|\Jzz=\mathrm{sgn}(\Jpm)B_4$ approaches $-5.5$ on the $\pi$-flux side and $+6.6$ on the $0$-flux side as $|\Jpm|\to0$, consistent with the perturbative prediction $B_4=6$ and distinct from the thermodynamic-limit value $B=15$.

\begin{figure*}[t]
\includegraphics[width=0.98\textwidth]{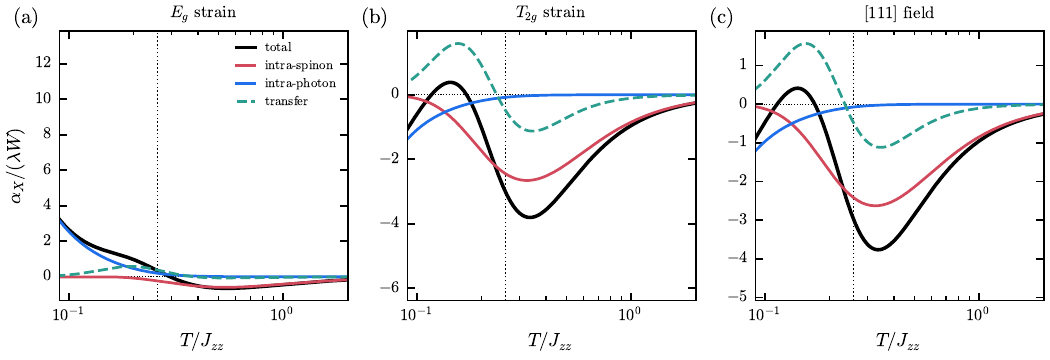}
\caption{Sector decomposition of the response [Eq.~\eqref{eq:sectorsplit}] at
$\Jpm=-0.10$ and $x=\lambda^2/|J_\pm|J_{zz}=0.014$, shown per unit field power for (a)~$E_g$ strain,
(b)~$T_{2g}$ strain, and (c)~a $[111]$ magnetic field. At the spinon anomaly
$T_{\rm sp}$ (dotted vertical line), the response is carried predominantly by
the intra-spinon term, i.e. the conditional covariance between the
polarizability and energy of the charged eigenstates. The transfer term becomes
important where thermal weight first leaves the ice-descended levels, while the
intra-ring contribution is negligible across the spinon window.}
\label{fig:sector}
\end{figure*}

\section{The spinon window}
\label{sec:spinon}

The charged sector of quantum spin ice has been studied extensively using gauge mean-field theory and related approaches~\cite{Savary2013,Chen2016,Chen2017,Udagawa2019,Desrochers2023,Desrochers2024,Desrochers2024FiniteT}, and its excitations are now being resolved spectroscopically~\cite{Gao2025Photons,Gao2026Demarcation}. Here we establish the two properties of the spinon-scale thermal response used in the main text: first, its dominant intra-spinon contribution has a definite sign that is opposite to the ring-sector response in the $\pi$-flux regime; second, its amplitude is strongly probe dependent because the same-sublattice and cross-sublattice pieces of the weak-field kernel interfere differently in the uniform and zero-sum channels.

\subsection{Sector decomposition of the response}
\label{sec:spinon:sector}

For diagnostic purposes, we partition the exact spectrum into the lowest $90$ levels, denoted the ring sector, and the remaining $65\,446$ levels, denoted the spinon sector. This is purely a bookkeeping decomposition of the same exact spectrum; no projection is introduced into the Hamiltonian. To keep $\sigma$ reserved for sublattice indices, we use $s\in\{\mathrm{ring},\mathrm{sp}\}$ as the sector label. Let $P_s(T)$ be the total Boltzmann weight of sector $s$, and let $\bar X_s$, $\bar E_s$, and $\Cov_s$ denote conditional means and covariances within that sector.

The covariance entering Eq.~\eqref{eq:maxwellSM} then obeys the exact law of total covariance,
\begin{equation}
  \Cov(\Xhat,H)=
  \underbrace{P_{\mathrm{ring}}\Cov_{\mathrm{ring}}}_{\text{intra-ring}}
  +\underbrace{P_{\rm sp}\Cov_{\rm sp}}_{\text{intra-spinon}}
  +\underbrace{P_{\mathrm{ring}}P_{\rm sp}
   \big(\bar X_{\rm sp}-\bar X_{\mathrm{ring}}\big)
   \big(\bar E_{\rm sp}-\bar E_{\mathrm{ring}}\big)}_{\text{transfer}} .
  \label{eq:sectorsplit}
\end{equation}
The first two terms measure correlations generated within the two sectors. The transfer term instead reflects redistribution of thermal weight between sectors whose mean energies and mean polarizations differ.

Figure~\ref{fig:sector} shows the three contributions at $\Jpm=-0.10$. At the spinon anomaly, $T_{\rm sp}=0.259$, the two zero-sum probes are overwhelmingly dominated by the intra-spinon covariance. Thus the spinon feature in $\gamma_X$ is primarily the covariance between the field-induced polarizability of charged eigenstates and their energy.

Per unit field power, the intra-spinon contributions at $T_{\rm sp}$ are $-0.24$ for $E_g$, $-2.46$ for $T_{2g}$, and $-2.43$ for the $[111]$ field. The uniform channel is therefore suppressed by roughly an order of magnitude, while the two zero-sum channels agree at the percent level, as required by Eq.~\eqref{eq:Reig}. In the uniform channel this suppression is strong enough that the transfer contribution can dominate the total response and reverse its sign. That sign change is therefore a consequence of the screened intra-spinon signal rather than an independent spinon mechanism.

\subsection{The spinon kernel and channel screening}
\label{sec:spinon:flavor}

Because the spinon anomaly is dominated by the intra-spinon covariance, it can be analyzed using the same two-component sublattice kernel introduced in Sec.~\ref{sec:pt:flavor}. We denote its spinon-sector components by $\chi_d^{\rm sp}$ and $\chi_o^{\rm sp}$. Equation~\eqref{eq:Reig} then gives
\begin{equation}
  \text{uniform:}\quad \chi_d^{\rm sp}+3\chi_o^{\rm sp},
  \qquad
  \text{zero sum:}\quad \chi_d^{\rm sp}-\chi_o^{\rm sp}.
\end{equation}

\begin{figure*}[t]
\includegraphics[width=0.98\textwidth]{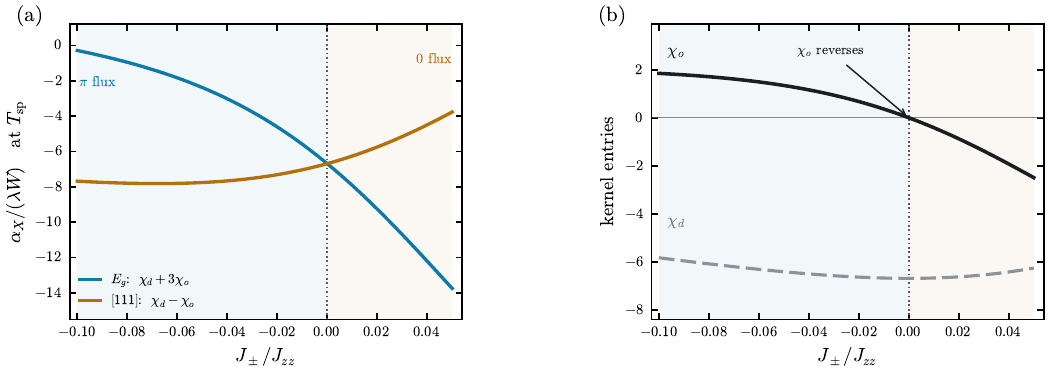}
\caption{Origin of the channel selectivity of the spinon anomaly.
Single-tetrahedron weak-field kernels evaluated at $T_{\rm sp}$.
(a)~The uniform combination $\chi_d+3\chi_o$, measured by $E_g$ strain, and
the zero-sum combination $\chi_d-\chi_o$, measured by a $[111]$ field, shown
per unit field power. The two cross at $\Jpm=0$ and exchange their relative
strengths: the uniform response is suppressed on the $\pi$-flux side, whereas
the field response is suppressed on the $0$-flux side.
(b)~The microscopic origin of this interchange. The diagonal component
$\chi_d$ is negative and only weakly dependent on $\Jpm$, while the
off-diagonal component $\chi_o$ vanishes at $\Jpm=0$ and reverses sign with
$\Jpm$. The resulting constructive or destructive interference determines
which laboratory channel is screened.}
\label{fig:spinon}
\end{figure*}

The two terms have different microscopic origins. The diagonal contribution $\chi_d^{\rm sp}$ describes the local transverse polarizability of a sublattice: $\Xhat_\mu$ tilts spins on sublattice $\mu$ away from their Ising axes against the longitudinal exchange environment. This process exists already at $\Jpm=0$, and Fig.~\ref{fig:spinon}(b) shows that it depends only weakly on $\Jpm$.

The off-diagonal contribution $\chi_o^{\rm sp}$ correlates transverse tilts on different sublattices and therefore requires transverse exchange. It vanishes at $\Jpm=0$ and changes sign with $\Jpm$. The flux sector thus determines whether transverse polarizations on neighboring sublattices interfere constructively or destructively in a given laboratory channel.

This sign reversal explains the channel selectivity in Fig.~\ref{fig:spinon}(a). Since $\chi_d^{\rm sp}<0$, the $\pi$-flux side has $\chi_o^{\rm sp}>0$, which partially cancels the diagonal response in the uniform combination $\chi_d^{\rm sp}+3\chi_o^{\rm sp}$ and suppresses the $E_g$ anomaly. On the $0$-flux side, $\chi_o^{\rm sp}<0$: it reinforces the uniform response but partially cancels the zero-sum combination $\chi_d^{\rm sp}-\chi_o^{\rm sp}$, suppressing the magnetic-field channel instead. The weak channel at the spinon scale is therefore itself a diagnostic of the flux sector.

\section{Observables}
\label{sec:obs}

\subsection{Strain channels and dilatometric coefficients}
\label{sec:obs:strain}

Under the cubic point group, the symmetric strain tensor $\varepsilon_{ij}$ decomposes into an $A_{1g}$ volume singlet, an $E_g$ doublet, and a $T_{2g}$ triplet~\cite{Dresselhaus2008},
\begin{align}
  \varepsilon_{A_{1g}}&=\tfrac{1}{\sqrt3}(\varepsilon_{xx}+\varepsilon_{yy}+\varepsilon_{zz}),\\
  \varepsilon_{E_g}&=\Big\{\tfrac{1}{\sqrt6}(2\varepsilon_{zz}-\varepsilon_{xx}-\varepsilon_{yy}),\
                 \tfrac{1}{\sqrt2}(\varepsilon_{xx}-\varepsilon_{yy})\Big\},
                 \label{eq:strainirrep}\\
  \varepsilon_{T_{2g}}&=\{\varepsilon_{yz},\,\varepsilon_{zx},\,\varepsilon_{xy}\}.
\end{align}
For a measurement axis $\hat{\bm\ell}$, the linear thermal-expansion coefficient is $\alpha(\hat{\bm\ell})=\sum_{ij}\alpha_{ij}\ell_i\ell_j$, with $\alpha_{ij}=d\varepsilon_{ij}/dT$~\cite{Patri2019Unveiling}. Along high-symmetry directions,
\begin{align}
    \alpha_{[100]}&=\alpha_{xx}, \\
    \alpha_{[110]}&=\tfrac12(\alpha_{xx}+\alpha_{yy})+\alpha_{xy},\\
    \alpha_{[1\bar10]}&=\tfrac12(\alpha_{xx}+\alpha_{yy})-\alpha_{xy},
    \label{eq:alphadir}  
\end{align}
and therefore
\begin{align}
  \alpha_{E_g}^{(1)}&\propto\alpha_{xx}-\alpha_{yy}=\alpha_{[100]}-\alpha_{[010]},\label{eq:egmeas}\\
  \alpha_{E_g}^{(2)}&\propto2\alpha_{zz}-\alpha_{xx}-\alpha_{yy}
     =2\alpha_{[001]}-\alpha_{[100]}-\alpha_{[010]},
\end{align}
while $\alpha_{xy}=\tfrac12(\alpha_{[110]}-\alpha_{[1\bar10]})$ isolates a $T_{2g}$ component. Taking differences between symmetry-related directions removes the $A_{1g}$ volume contribution and directly projects the measured expansion onto the desired irreducible representation. Equation~\eqref{eq:egmeas} is the relation used in the main text.

\subsection{Magnetoelastic coupling and the fixed-stress thermal-expansion response}
\label{sec:obs:magnetoelastic}

For a non-Kramers doublet, the transverse pseudospin components transform as electric quadrupoles and therefore couple linearly to strain~\cite{OnodaTanaka2010,OnodaTanaka2011,Onimaru2016,Simon2022,Seth2022,Tang2023,Tang2024,Chung2025}. The same microscopic coupling allows random strain to act as a random transverse field in $\mathrm{Pr_2Zr_2O_7}$~\cite{Wen2017,SavaryBalents2017,Benton2018,Luo2025Pr,Marinho2026}; here we instead use its spatially uniform, symmetry-resolved component as a controlled thermodynamic probe.

For one component of a symmetry channel $\mathcal R\in\{E_g,T_{2g}\}$, the magnetoelastic coupling can be written
\begin{equation}
  H_{\rm me}=-g_{\mathcal R}\varepsilon_{\mathcal R}O_{\mathcal R}.
  \label{eq:HmeSM}
\end{equation}
Comparison with $H(\lambda)=H_0-\lambda O_{\mathcal R}$ identifies the generalized field used throughout the microscopic calculation as the energy-like quantity
\begin{equation}
  \lambda=g_{\mathcal R}\varepsilon_{\mathcal R}.
  \label{eq:lambdastrainsm}
\end{equation}
Thus the fixed-$\lambda$ covariance response computed in the main text is the electronic response at fixed strain. A dilatometry experiment instead measures the equilibrium strain at fixed applied stress. To connect the two ensembles, introduce the thermodynamic potential
\begin{equation}
  \Phi(T,\varepsilon_{\mathcal R};\sigma_{\mathcal R})
  =F_{\rm spin}\!\left(T,\lambda=g_{\mathcal R}\varepsilon_{\mathcal R}\right)
  +\frac{c_{\mathcal R}}{2}\varepsilon_{\mathcal R}^2
  -\sigma_{\mathcal R}\varepsilon_{\mathcal R},
  \label{eq:PhiStressSM}
\end{equation}
where $c_{\mathcal R}$ is the bare elastic stiffness and $\sigma_{\mathcal R}$ is the externally applied symmetry-resolved stress. Since $\partial_\lambda F_{\rm spin}=-\expv{O_{\mathcal R}}$, minimization with respect to strain gives
\begin{equation}
  c_{\mathcal R}\varepsilon_{\mathcal R}^*
  =\sigma_{\mathcal R}+g_{\mathcal R}\expv{O_{\mathcal R}}.
  \label{eq:strainstationarySM}
\end{equation}
The weak stress supplies the finite symmetry-breaking strain, while Eq.~\eqref{eq:lambdastrainsm} shows explicitly how this strain generates the transverse field entering the ring-exchange calculation.

Define the fixed-strain thermal response and the static susceptibility
\begin{equation}
  \gamma_{\mathcal R}
  \equiv\left.\frac{\partial\expv{O_{\mathcal R}}}{\partial T}\right|_{\lambda},
  \qquad
  \chi_{\mathcal R}
  \equiv\left.\frac{\partial\expv{O_{\mathcal R}}}{\partial\lambda}\right|_{T}.
  \label{eq:gammachiSM}
\end{equation}
Differentiating Eq.~\eqref{eq:strainstationarySM} with respect to temperature at fixed applied stress, and using $d\lambda=g_{\mathcal R}d\varepsilon_{\mathcal R}^*$, gives
\begin{equation}
  c_{\mathcal R}\alpha_{\mathcal R}
  =g_{\mathcal R}\left(\gamma_{\mathcal R}
  +g_{\mathcal R}\chi_{\mathcal R}\alpha_{\mathcal R}\right),
\end{equation}
where $\alpha_{\mathcal R}=(\partial_T\varepsilon_{\mathcal R}^*)_{\sigma_{\mathcal R}}$. Hence
\begin{equation}
  \boxed{
  \alpha_{\mathcal R}
  =\frac{g_{\mathcal R}}
  {c_{\mathcal R}-g_{\mathcal R}^2\chi_{\mathcal R}}
  \gamma_{\mathcal R}}
  \equiv
  \frac{g_{\mathcal R}}{c_{\mathcal R}^{\rm eff}}\gamma_{\mathcal R}.
  \label{eq:melock}
\end{equation}
Equation~\eqref{eq:melock} is the exact fixed-stress relation within the harmonic elastic model. Neglecting magnetoelastic feedback, $g_{\mathcal R}^2\chi_{\mathcal R}\ll c_{\mathcal R}$, it reduces to the main-text expression $\alpha_{\mathcal R}\simeq(g_{\mathcal R}/c_{\mathcal R})\gamma_{\mathcal R}$. The feedback therefore changes only the elastic prefactor; provided $c_{\mathcal R}^{\rm eff}(T)$ is smooth across the narrow ring-exchange anomaly, it does not alter the location or the flux-dependent sign of the microscopic ring-sector response.

The main text uses $O_{E_g}=\sum_i(\sqrt3S^x_i-S^y_i)$, whose sublattice amplitudes $\bm p^{E_g}=e^{i\pi/6}(1,1,1,1)$ are uniform; the $T_{2g}$ channel replaces the uniform site form factor with $\bm p^{T_{2g}}$ of Eq.~\eqref{eq:strainvectorsSM}. Both strains are uniform in real space ($\bm q=0$), but they occupy different sublattice channels: $E_g$ is uniform while $T_{2g}$ is zero sum. Sections~\ref{sec:pt:flavor} and \ref{sec:spinon:flavor} show that this sublattice distinction determines their relative weak-field response.

\subsection{Absence of a direct nuclear Schottky term}
\label{sec:obs:nuclear}
Let the spin-plus-nuclear Hamiltonian be
\begin{equation}
  H(\lambda)=H_{\rm el}(\lambda)+H_{\rm hf},
  \label{eq:hfstrainindependent}
\end{equation}
where we assume that the hyperfine Hamiltonian has no explicit dependence on the symmetry-breaking strain, $\partial_\lambda H_{\rm hf}=0$. The generalized force conjugate to $\lambda$ is then
\begin{equation}
  -\partial_\lambda H
  =-\partial_\lambda H_{\rm el}
  =O_{\mathcal R},
  \label{eq:nohfX}
\end{equation}
so the nuclear Hamiltonian produces no independent generalized-force or stress term. Equivalently,
\begin{equation}
  \gamma_{\mathcal R}
  =\frac{1}{T^2}\Cov_T(O_{\mathcal R},H_{\rm el})
   +\frac{1}{T^2}\Cov_T(O_{\mathcal R},H_{\rm hf}).
  \label{eq:hfcovSM}
\end{equation}
There is therefore no additive contribution proportional to $\Var(H_{\rm hf})/T^2$, i.e. no direct nuclear Schottky heat-capacity background in the symmetry-resolved thermal-expansion channel. Hyperfine coupling can still renormalize the electronic response indirectly through the equilibrium density matrix and the mixed covariance in the second term of Eq.~\eqref{eq:hfcovSM}. This is the precise sense in which the nuclear Schottky background is absent from the probe proposed in the main text.

\end{document}